\documentclass[usenatbib]{mnras}

\usepackage{newtxtext,newtxmath}
\usepackage[T1]{fontenc}

\usepackage{graphicx}	
\usepackage{amsmath}
\usepackage{soul,xcolor}
\DeclareMathAlphabet\mathbfcal{OMS}{cmsy}{b}{n}
\usepackage{bbold}

\usepackage{stmaryrd}
\usepackage{trimclip}

\makeatletter
\DeclareRobustCommand{\shortto}{%
  \mathrel{\mathpalette\short@to\relax}%
}

\newcommand{\short@to}[2]{%
  \mkern2mu
  \clipbox{{.5\width} 0 0 0}{$\m@th#1\vphantom{+}{\shortrightarrow}$}%
  }
\makeatother

\DeclareMathAlphabet\mathbfcal{OMS}{cmsy}{b}{n}

\setstcolor{red}

\title[Magnetically driven coupling in relativistic RMS]{Magnetically driven coupling in relativistic radiation-mediated shocks}

\author[Mahlmann et al.]{
J. F. Mahlmann,$^{1}$\thanks{E-mail: mahlmann@princeton.edu}
A. Vanthieghem,$^{1,2}$
A. A. Philippov,$^{3}$
A. Levinson,$^{4}$
E. Nakar,$^{4}$
F. Fiuza$^{5}$
\\
$^{1}$ Department of Astrophysical Sciences, Peyton Hall, Princeton University, Princeton, NJ 08544, USA\\
$^{2}$ International Research Collaboration Center, National Institutes of Natural Sciences, Tokyo, 105-0001, Japan\\
$^{3}$ Department of Physics, University of Maryland, College Park, MD 20742, USA\\
$^{4}$ The Raymond and Beverly Sackler School of Physics and Astronomy, Tel Aviv University, Tel Aviv 69978, Israel\\
$^{5}$ High Energy Density Science Division, SLAC National Accelerator Laboratory, Menlo Park, CA 94025, USA\\
}

\date{Accepted 2023 January 6. Received 2023 January 5; in original form 2022 November 9}

\pubyear{2022}

\begin{document}

\label{firstpage}
\pagerange{\pageref{firstpage}--\pageref{lastpage}}
\maketitle

\begin{abstract}
The radiation drag in photon-rich environments of cosmic explosions can seed kinetic instabilities by inducing velocity spreads between relativistically streaming plasma components. Such microturbulence is likely imprinted on the breakout signals of radiation-mediated shocks. However, large-scale, transverse magnetic fields in the deceleration region of the shock transition can suppress the dominant kinetic instabilities by preventing the development of velocity separations between electron-positron pairs and a heavy ion species. We use a one-dimensional (1D) five-fluid radiative transfer code to generate self-consistent profiles of the radiation drag force and plasma composition in the deceleration region. For increasing magnetization, our models predict rapidly growing pair multiplicities and a substantial radiative drag developing self-similarly throughout the deceleration region. We extract the critical magnetization parameter $\sigma_{c}$, determining the limiting magnetic field strength at which a three-species plasma can develop kinetic instabilities before reaching the isotropized downstream. For a relativistic, single ion plasma drifting with $\gamma_{u} = 10$ in the upstream of a relativistic radiation-mediated shock, we find the threshold $\sigma_{c}\approx 10^{-7}$ for the onset of microturbulence. Suppression of plasma instabilities in the case of multi-ion composition would likely require much higher values of $\sigma_{c}$. Identifying high-energy signatures of microturbulence in shock breakout signals and combining them with the magnetization limits provided in this work will allow a deeper understanding of the magnetic environment of cosmic explosions like supernovae, gamma-ray bursts, and neutron star binary mergers. 
\end{abstract}




\section{Introduction}
\label{sec:introduction}

Relativistic radiation-mediated shocks (RMS) can generate the first bright burst of radiation observed from powerful cosmic explosions. Such transients include the early phases of supernovae \citep[e.g.,][]{Soderberg2008,Chevalier2008,Xiang2019}, the prompt emission of gamma-ray bursts \citep[GRBs; e.g.,][]{Page2007,Abdo2009}, and short GRBs from neutron star binary mergers \citep{Goldstein2017}. Systematic observation of such fugitive flashes remains a challenge even with current wide-view surveys \citep{Bayless2022}. Still, when captured and paired with theoretical models of radiative processes \citep{Levinson2020}, they provide a rich insight into the precursor environment. This paper examines how the magnetic field strength, varying significantly between different progenitor systems, impacts the micro-physics on kinetic scales of radiation-mediated shocks.

In optically thick media, radiative, quantum-electrodynamics (QED) interactions drive the dissipation that injects the energy stored in collective plasma motions into a reservoir of highly energetic photons \citep{Lundman2018}. It is this radiation loaded around the shock transition layer that is released promptly during the so-called shock breakout \citep{nakar2012,Waxman2017}. Compared to collisionless shocks, mediated by the collective plasma dynamics on kinetic scales associated with the particle Larmor radius, RMS are generally much wider and extend over several photon mean free paths \citep{levinson2012}. Plasma kinetic effects can then act locally, well-separated from the global RMS size. Instabilities can develop on small scales under certain conditions, such as the microturbulence examined in an earlier work by \citet{Vanthieghem2022}. One trigger for growing wave modes is the relative drift velocity of different particle species induced by the different slow-down rates in the deceleration region \citep{levinson2020b}. Plasma components decouple and develop a relative drift velocity due to an imbalance between the radiative force, acting equally on lepton species, and the restoring electric field that accelerates electrons and positrons into different directions. Growing unstable modes can seed kinetic-scale electromagnetic fields. This microturbulence in the radiation-rich downstream is likely to impact signals observed during the shock breakout.

In magnetized environments, a transverse magnetic field generates oscillatory velocity drifts that couple the various plasma components. While a radiative force drives small velocity separations, the coupled plasma will partially compensate for the deceleration of light particles (i.e., leptons) by growing an intermittent longitudinal electric field. This magnetization-induced force decelerates the heavy ions while re-accelerating the leptons. In this state, the deceleration profiles resemble those of a single fluid. The strength of the magnetic fields in the astrophysical environments where RMS are expected to operate is not well constrained. It likely varies widely between different scenarios of cosmic explosions, but again within the different systems of the same explosion mechanism. In this paper, we, therefore, study the impact of transverse magnetic fields independent from the progenitor system to determine \emph{at which critical magnetization a radiation-mediated shock will likely develop kinetic instabilities}. An observational signature of microturbulence during the shock breakout, or the lack thereof, would then allow for setting limits on the magnetic properties of the shock generation site, and a deeper understanding of its immediate astrophysical environment. 

This paper extends our previous work on the growth of micro-instabilities in an unmagnetized environment \citep{Vanthieghem2022}. Section~\ref{sec:1DModeling} reviews the numerical methods used for a (self-consistent) radiative transfer modeling of RMS deceleration regions. We present the employed units and normalization (Section~\ref{sec:units}), as well as the 1D multi-fluid radiative transfer system of differential equations (Section~\ref{sec:1DRadiative}). We numerically evaluate the magnetic coupling of different species in Section~\ref{sec:1DEquilibrium}, both for equilibrium configurations with a constant radiative force (\ref{sec:couplingomega}), and consistently modeled radiative transfer (\ref{sec:pairdeceleration}). Section~\ref{sec:thresholds} contains the main findings of this paper and gives specific magnetization thresholds for the strong coupling of plasma components in the deceleration region. We give a summary of our results and their implications in Section~\ref{sec:discussion}. A detailed review of the linear plasma response is given in Appendix~\ref{app:linearanalysis}, the numerical implementation is presented in Appendix~\ref{app:imexstepping} and calibrated in Appendix~\ref{app:calibrationB0}, Appendix~\ref{eq:FradNorm} reviews local plasma scalings and unit systems.

\section{Methodology}
\label{sec:1DModeling}

\subsection{Units and normalization}
\label{sec:units}

The physics described in this paper covers largely different time and length scales. The coordinates ($ct$ and $x$) are scaled to the Thomson length in the \emph{shock-front-frame} $\lambda = 1/(\sigma_{\rm T}\bar{n}_{u} \gamma_{u})$, where $\sigma_{\rm T}$ is the Thomson cross section, $\bar{n}_{u}$ is the upstream density measured in the \emph{upstream baryon frame} (denoted by a bar), and $\gamma_{u}$ is the upstream baryon Lorentz factor in the shock-front frame. Densities in the \emph{shock-front-frame} are rescaled to the upstream baryon density as $\tilde{n}=\bar{n}\gamma/(\bar{n}_{u}\gamma_{u})$. Electromagnetic fields ${E}_x$, ${E}_y$, and ${B}_z$ are normalized by the fiducial electric field (denoted by a tilde)
\begin{align}
    E_0 = \frac{m_e c^2}{\lambda e} \approx 1.13\times 10^{-6} \;\gamma_{u}\;\left(\frac{n_{u}}{10^{15}\text{cm}^{-3}}\right)\text{G} \,.
\end{align}
Our numerical scheme employs a normalization that isolates the dimensionless scale-separation factor, $\chi_0$:
\begin{align}
    \chi_0=\frac{4\pi e}{\sigma_{\rm T} E_0}=8.0\times 10^{21} \;\gamma_{u}^{-1}\;\left(\frac{10^{15}\text{cm}^{-3}}{n_{u}}\right) \,.
    \label{eq:scaleseparation}
\end{align}
This factor is proportional to the second power of the ratio between the radiative scale $\lambda$ and the electron skin depth $\bar{d}_e$. The presented multi-fluid plasma models cannot resolve realistic scale separations. Therefore, we introduce $\chi$ as a free parameter chosen according to the limits of the respective numerical scheme and extrapolate converged results to realistic scales (where $\chi\rightarrow\chi_0$). The key parameters that characterize the plasma in terms of the scale-separation factor $\chi$ are the magnetization, the Larmor radius, the plasma skin depth, and the radiative length. The magnetization of the \emph{upstream} with mass ratio $\mu = m_e / m_{p}$ is
\begin{align}
\bar{\sigma}_{u}=\frac{B_{u}^2}{4\pi  m_{p} c^2 \bar{n}_{u} \gamma_{u}^2}&= \frac{\tilde{B}_{u}^2 E_0^2}{4\pi  m_{p} c^2 \bar{n}_{u} \gamma_{u}^2} = \frac{\mu \tilde{B}_{u}^2}{\chi\gamma_{u}},
\label{eq:magnetization}
\end{align}
Here, $B_{u}$ is the transverse (along the $z$-direction) magnetic field strength measured in the \emph{shock-front frame}. The Larmor radius in the \emph{frame of the upstream} plasma is
\begin{align}
    \begin{split}
      \bar{\rho}&=\gamma_{u}\frac{m_e c^2}{e B_{u}}=\frac{m_e c^2}{e E_0}\left(\frac{\gamma_{u}\mu }{\chi\bar{\sigma}_{u}}\right)^{1/2}=\mu^{1/2}\left(\frac{ \gamma_{u}}{\chi\bar{\sigma}_{u}}\right)^{1/2} \lambda \,.
    \label{eq:larmorradius}
    \end{split}
\end{align}
We note that the product $\chi\bar{\sigma}_u$ fully determines the Larmor radius for given $\gamma_{u}$ and $\bar{n}_{u}$. The ion plasma skin depth is:
\begin{align}
    \begin{split}
        \bar{d}&=\frac{c}{\bar{\omega}_{pu}}=\mu^{-1/2}\left(\frac{m_e c^2}{4\pi e^2 \bar{n}_{u}}\right)^{1/2}=\mu^{-1/2}\left(\frac{\gamma_{u}}{\chi}\right)^{1/2}\lambda \,.
        \label{eq:skindepth}
    \end{split}
\end{align}
Combining Equations~(\ref{eq:larmorradius}) and~(\ref{eq:skindepth}) yields $\bar{d}/\bar{\rho}= \bar{\sigma}_{u}^{1/2}\mu^{-1}$. The presented simulations rely on a careful evaluation of the transition to a realistic scale separation. Increasing $\chi$ reduces the plasma skin depth or, vice versa raises the plasma frequency. Thus, and despite using implicit methods (Appendix~\ref{app:imexstepping}), the numerical integration for realistically large values of $\chi\rightarrow\chi_0$ becomes prohibitive due to the high resolution required to resolve the smallest scales.

\subsection{1D-multi-fluid radiative transfer}
\label{sec:1DRadiative}

In this section, we discuss a 1D radiative transfer model formulated as a system of equations coupling the flow of five fluids with the dynamics of electromagnetic fields. The fluids encode the collective motion of individual species ($s$), namely protons ($p$), electrons ($-$), and positrons ($+$), as well as two photon beams flowing towards the upstream ($\gamma\shortto u$) and towards the downstream ($\gamma\shortto d$). The continuity equations for the different species are then given by:
\begin{align}
    & \partial_{\tilde{t}}\tilde{n}_{p}+\partial_{\tilde{x}} \left( \tilde{n}_{p} \beta^x_{p} \right) =0\label{eq:momentumions}\,,\\
    & \partial_{\tilde{t}}\tilde{n}_\pm+\partial_{\tilde{x}} \left(\tilde{n}_\pm\beta^x_\pm\right) = 2\tilde{\sigma}_{\gamma\gamma}\tilde{n}_{\gamma\shortto d}\tilde{n}_{\gamma\shortto u}\label{eq:momentumpairs}\,,\\
    \begin{split}
    &    \partial_{\tilde{t}} \tilde{n}_{\gamma\shortto u}+\partial_{\tilde{x}} \left(\tilde{n}_{\gamma\shortto u}\beta^x_{\gamma\shortto u}\right) = \\ 
    &\qquad\qquad -2\left(\tilde{\sigma}_+ \tilde{n}_+ + \tilde{\sigma}_-  \tilde{n}_- + \tilde{\sigma}_{\gamma\gamma} \tilde{n}_{\gamma\shortto d}\right) \tilde{n}_{\gamma\shortto u}\,,
    \end{split}\label{eq:photonmoments1}
    \\
    \begin{split}
    &    \partial_{\tilde{t}} \tilde{n}_{\gamma\shortto d}+\partial_{\tilde{x}} \left(\tilde{n}_{\gamma\shortto d}\beta^x_{\gamma\shortto d}\right) = \\ 
    &\qquad\qquad 2\left(\tilde{\sigma}_+ \tilde{n}_+ + \tilde{\sigma}_-  \tilde{n}_- - \tilde{\sigma}_{\gamma\gamma} \tilde{n}_{\gamma\shortto d}\right) \tilde{n}_{\gamma\shortto u}\,.
    \end{split}\label{eq:photonmoments2}
\end{align}
Following the derivations by \citet{levinson2020b}, we include the source term of Equation~(\ref{eq:momentumpairs}) to account for pair production with a respective sink in the photon beams (Equations~\ref{eq:photonmoments1} and~\ref{eq:photonmoments2}). Here, $\sigma_{\gamma\gamma}$ is the pair production cross section. Terms proportional to $\sigma_\pm$, the Compton scattering cross section, balance the inverse Compton scattering of beam photons with a transition from ($\gamma\shortto u$) to ($\gamma\shortto d$). The total 4-current density is conserved, as one can see from building the sum over Equations~(\ref{eq:momentumions}) to~(\ref{eq:photonmoments2}). For the remainder of this paper, we assume highly beamed photons with $-\beta^x_{\gamma\shortto u}=\beta^x_{\gamma\shortto d}=1$, such that only the momentum equations for protons and pairs need to be integrated:
\begin{align}
    \left(\partial_{\tilde{t}}+\beta^x_{p}\partial_{\tilde{x}}\right) u^x_{p} &=\mu\left(\tilde{E}_x+\beta^y_{p} \tilde{B}_z\right)\label{eq:protonmomentum} \,,\\
    \left(\partial_{\tilde{t}}+\beta^x_\pm\partial_{\tilde{x}}\right)\left(h_\pm u^x_\pm\right)&=\pm\left(\tilde{E}_x+\beta^y_\pm \tilde{B}_z\right)-\tilde{\mathcal{F}}_{\rm rad}\label{eq:upmmomentum} \,,\\
    \left(\partial_{\tilde{t}}+\beta^x_{p}\partial_{\tilde{x}}\right) u^y_{p}&=\mu\left(\tilde{E}_y-\beta^x_{p} \tilde{B}_z\right) \,,\\
    \left(\partial_{\tilde{t}}+\beta^x_\pm\partial_{\tilde{x}}\right)\left(h_\pm u^y_\pm\right)&=\pm\left(\tilde{E}_y-\beta^x_\pm \tilde{B}_z\right)\label{eq:pairymomentum2} \,.
\end{align}
Here, $\tilde{\mathcal{F}}_{\rm rad}$ denotes the Compton scattering drag in the Klein-Nishina (KN) regime, or radiative force, which can be consistently modeled as $\tilde{\mathcal{F}}_{\rm rad}=2\tilde{\sigma}_\pm h_\pm u^x_\pm \tilde{n}_{\gamma\shortto u}$. We generally assume that leptons have the same temperature and baryons are cold, such that $h_\pm = 1+4\hat{T}$. Following \citet{granot2018}, we approximate
\begin{align}
    \hat{T}=\eta\frac{\tilde{n}_{l}}{2+\tilde{n}_{l}}\frac{\gamma_+ + \gamma_-}{2} \,.
\end{align}
The density of quanta, $\tilde{n}_{l}$, combines the density of pairs created inside the shock with the contribution from back-scattered photons,
\begin{align}
    \tilde{n}_{l}=\tilde{n}_+ - \left[\tilde{n}_{p}\left(\tilde{x}=0\right)-\tilde{n}_- \right] + \tilde{n}_{\gamma\shortto d} \,,
\end{align}
and we adopt $\eta = 0.45$. With these definitions, the scattering cross sections (normalized to $\sigma_{\rm T}$) are
\begin{align}
    \tilde{\sigma}_\pm = \frac{3}{8}\frac{\ln\left[2\gamma_\pm\left(1+a\hat{T}\right)\right]}{\gamma_\pm\left(1+a\hat{T}\right)}\,,\qquad\qquad\tilde{\sigma}_{\gamma\gamma}=\frac{\tilde{\sigma}_+ + \tilde{\sigma}_-}{2}.
\end{align}
Finally, the continuity equations (\ref{eq:momentumions}) to (\ref{eq:photonmoments2}) are coupled to the momentum equations (\ref{eq:protonmomentum}) to (\ref{eq:pairymomentum2}) via:
\begin{align}
    \partial_{\tilde{t}}\tilde{E}_x&=-\chi\left[\tilde{n}_{p} \beta^x_{p}+\tilde{n}_+\beta^x_+ -\tilde{n}_-\beta^x_-\right]\label{eq:couplingIIIa} \,,\\
    \partial_{\tilde{t}}\tilde{E}_y&=-\chi\left[\tilde{n}_{p} \beta^y_{p}+\tilde{n}_+\beta^y_+ -\tilde{n}_-\beta^y_-\right]-\partial_{\tilde x}\tilde{B}_z \,,\\
    \partial_{\tilde{t}}\tilde{B}_z&=-\partial_{\tilde x}\tilde{E}_y\label{eq:couplingIII} \,.
\end{align}
We integrate this system of equations with the method described and profiled in appendices~\ref{app:imexstepping} and~\ref{app:calibrationB0}. For studying multi-fluid equilibria in the deceleration region independent of the downstream (shock) conditions, it is convenient to replace Equation~(\ref{eq:photonmoments1}) for $\beta^x_{\gamma\shortto u}=-1$ by \citep[cf.][]{levinson2020b} 
\begin{align}
\begin{split}
    & \partial_{\tilde{t}} \tilde{n}_{\gamma\shortto u}+\partial_{\tilde{x}} \tilde{n}_{\gamma\shortto u} = 2\left(\tilde{\sigma}_+ \tilde{n}_+ + \tilde{\sigma}_-  \tilde{n}_- + \tilde{\sigma}_{\gamma\gamma} \tilde{n}_{\gamma\shortto d}\right) \tilde{n}_{\gamma\shortto u} \,.
    \end{split}
    \label{eq:trickeq}
\end{align}
This adaptation is a numerical trick that solves the system of Equations~(\ref{eq:momentumions}) to~(\ref{eq:couplingIII}) by driving it to an equilibrium state with $\partial_{\tilde{t}} \tilde{n}_{\gamma\shortto u}\rightarrow 0$, as it is expected in the shock front frame. We emphasize two aspects of this altered system of equations. First, the reformulation of Equation~(\ref{eq:trickeq}) allows us to use the time-dependent system of equations outlined above to find steady-state solutions of the deceleration profiles (time-independent). Second, these equilibrium profiles are decoupled from the physical conditions at the shock itself, most notably they do not include \emph{a priori} information about the photon beam $\tilde{n}_{\gamma\shortto u}$ injected downstream. However, it allows us to study the deceleration profiles of the different particle species and their dependence on upstream conditions (prescribed by $\gamma_{u}$ and $\bar{\sigma}_{u}$).

In the following sections, we evolve the full five-fluid system in time, starting at $\Tilde{t}=0$. We choose the initial and upstream boundary conditions of $\gamma_{\pm}=\gamma_{p}=\gamma_{u}$ with $\gamma_{u}=10$, a pair multiplicity of $\mathcal{M}=1$, a realistic mass ratio of $1/\mu=1836$, as well as a seed radiation field of $\tilde{n}_{\gamma\shortto u}=0.01$ and $\tilde{n}_{\gamma\shortto d}=0$. These boundary conditions follow the setup described by \citet[][Section III. B.]{Levinson2020}. As we explore the effect of magnetization and scale separation, 
controlled by $\bar{\sigma}_{u}$ and $\chi$, the discrete mesh and domain extensions are stated separately for each numerical experiment. In the downstream, we employ zero gradient boundary conditions. In practice, we evolve each setup in time either until the equilibrium state is established in the whole domain or the numerical solution breaks down due to extreme density and velocity gradients in the deceleration profile (and, hence, unresolved plasma scales).

\vspace{11pt}

\section{Magnetized equilibrium configurations}
\label{sec:1DEquilibrium}

\begin{figure}
  \centering
  \includegraphics[width=0.45\textwidth]{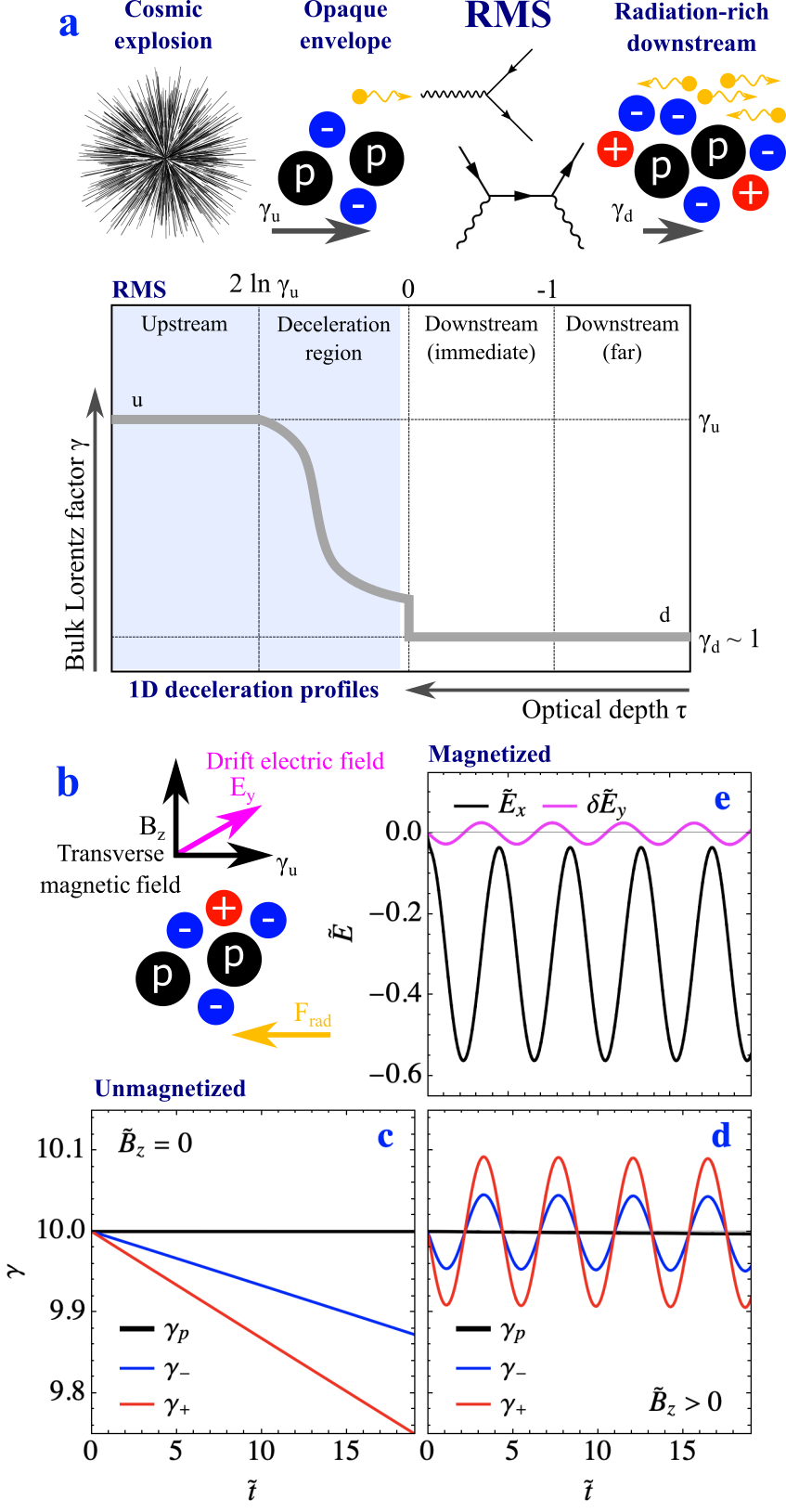}
  \vspace{-6pt}
  \caption{Schematic outline of typical (reversed) relativistic radiation-mediated shock profiles (RMS, panel a). Cartoons of particles indicate the respective plasma compositions (black: ions; blue: electrons; red: positrons; orange: photons). In (b) we show a cartoon of the employed setup. Panels (c) to (e) show solutions of deceleration profiles for periodic systems with a constant radiative force acting on lepton species \citep[see][Section~5]{Vanthieghem2022}. Without magnetic fields (panel c), the different particle species quickly decouple and develop a significant velocity separation \citep[see][]{levinson2020b}. In the magnetized case (panels d/e), leptons are coupled to the ions. Oscillations of the leptons are induced by a longitudinal electric field $\tilde{E}_x$ (panel e). Such fields effectively decelerate the coupled ions and can prevent the development of a large velocity spread.}
\label{fig:shematics}
\end{figure}

Figure~\ref{fig:shematics} (panel a) shows the astrophysical context of the photon-rich target environment. While lepton and ion species quickly decouple in the unmagnetized case \citep[seeding various plasma modes as discussed in][panel c]{Vanthieghem2022}, the addition of a transverse magnetic field efficiently couples the particles and prevents the growth of velocity drift (panels d/e). A self-consistent description of this coupling mechanism is the subject of this work.

One important auxiliary quantity for the interpretation of the deceleration profiles is the relative drift between ions and leptons measured in the \emph{local baryon frame} and denoted  $\bar{u}_\pm$. The respective boosts along $u_{p}^x$ can be written as 
\begin{align}
    \bar{u}_\pm=u_\pm^x \gamma_{p}^x-u_{p}^x \gamma_\pm^x \,,
    \label{eq:velocityspread}
\end{align}
where the Lorentz factors are associated with the bulk motion along the $x$-direction. The pair multiplicity $\mathcal{M}=\tilde{n}_+/\tilde{n}_{p}$ is the other relevant plasma characteristic that determines the strength of inter-species coupling and thereby affects the development of microturbulence in the deceleration region. Finally, we define the pair-loaded optical depth \citep[omitting Klein-Nishina effects, see][]{levinson2020b,granot2018}:
\begin{align}
    \tau^*=\int\tilde{n}_{l}\;\text{d}\tilde{x} \,.
\end{align}

\begin{figure}
  \centering
  \includegraphics[width=0.475\textwidth]{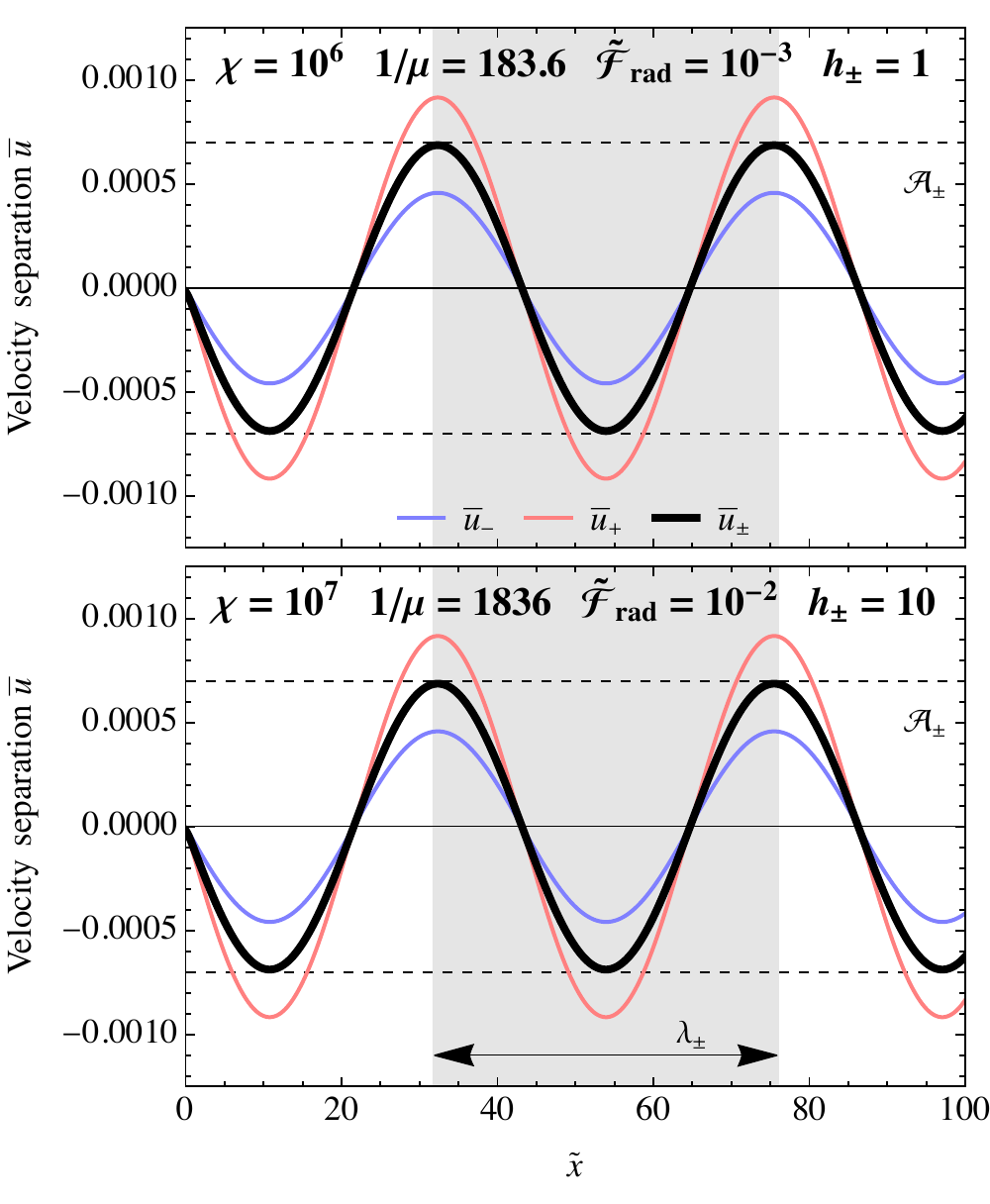}
  \vspace{-11pt}
  \caption{Comparison of the multi-fluid integration for $\tilde{\mathcal{F}}_{\rm rad}=\text{const.}$ in the cases of cold plasma ($h_\pm = 1$, top panel) and hot plasma ($h_\pm = 10$, bottom panel) with the re-scaling of parameters discussed in Section~\ref{sec:couplingomega}. The dynamics of inter-species coupling agree for both cases, and the cold-plasma estimates of Equations~(\ref{eq:couplinglength}) and~(\ref{eq:couplingamplitude}) are equally valid.}
\label{fig:ComparisonH}
\end{figure}

\subsection{Scales of lepton-ion coupling with constant radiative force}
\label{sec:couplingomega}

This section reviews the coupling of different species for a constant radiative force $\tilde{\mathcal{F}}_{\rm rad}$ used as a source term in Equation~(\ref{eq:upmmomentum}). Normalized to the upstream conditions $m_{p}\bar{\omega}_{pu}c$ one can write
\begin{align}
    \mathcal{F}_{\rm rad}=-\tilde{\mathcal{F}}_{\rm rad}\mu^{1/2}\chi^{-1/2}\gamma_{u}^{1/2}m_{p}\bar{\omega}_{pu}c\,.
    \label{eq:radforcenorm}
\end{align}
We present a linear analysis of the cold pair plasma response ($h_\pm=1$, $\hat{T}=0$) in Appendix~\ref{app:linearanalysis}. Equation~(\ref{eq:ApproxOmega}) suggests that in the far upstream, the scales of coupling between different species are well-separated from the plasma scales:
\begin{align}
    \frac{\bar{\omega}_\pm}{\bar{\omega}_{pu}}&=\gamma_{u}^{-2}\mu^{-1}\frac{\bar{\sigma}_{u}^{1/2}}{1+2 \mathcal{M}}\approx 10^{-2}\left(\frac{\gamma_{u}}{10}\right)^{-2} \mathcal{M}^{-1}\left(\frac{\bar{\sigma}_{u}}{10^{-6}}\right)^{1/2}\,.
    \label{eq:couplingomega}
\end{align}
Using Equation~(\ref{eq:skindepth}), the length scale of inter-species coupling written with local plasma properties in the shock-front frame is
\begin{align}
\begin{split}
    \lambda_\pm=\frac{2\pi}{{\omega}_\pm}&=2\pi\frac{\gamma^2}{\gamma_{u}}\left(\frac{\Tilde{B}_{u}}{\Tilde{B}_z}\right)\left(1+2\mathcal{M}\right)\bar{\rho}\,.
    \label{eq:couplinglength}
\end{split}
\end{align}
It is clear that commonly $\lambda_\pm\gg \rho$. The mean amplitude of the lepton oscillation around the slowly decelerating ions, as evaluated for far upstream conditions in \citet[][Section~5]{Vanthieghem2022}, is:
\begin{align}
\begin{split}
    \mathcal{A}_\pm&=\left(1+2\mathcal{M}\right)\tilde{\mathcal{F}}_{\rm rad}\;\frac{\bar{\rho}}{\lambda} \,.
    \label{eq:couplingamplitude}
\end{split}
\end{align}
The results displayed in Figure~\ref{fig:Validation} compare the linear estimates~(\ref{eq:couplingomega}) and~(\ref{eq:couplingamplitude}) to the direct integration of the effective three-fluid system. The good agreement between numerical results and analytic estimates validates the arguments we brought forward in \citet[][Section~5]{Vanthieghem2022}. As we will discuss throughout this section, the linear-theory estimate does not include the effects of finite lepton temperatures as well as the rapidly changing fluid dynamics during strong deceleration. In this section, we will address the first of these points, namely the dependence of the cold plasma scalings derived in Appendix~\ref{app:linearanalysis} on the temperature model introduced in Section~\ref{sec:1DRadiative}.

We examine the plasma response for a heated plasma (constant $h_\pm>1$, $\hat{T}>0$). Dividing the momentum equations~(\ref{eq:protonmomentum}) to~(\ref{eq:pairymomentum2}) by $h_\pm$ and adjusting the currents accordingly, a change in $h_\pm$ corresponds to re-scaling the cold-plasma characteristics as $\mu^h= h_\pm\mu$,  $\chi^h=\chi/h_\pm$, and $\tilde{\mathcal{F}}_{\rm rad}^h= \tilde{\mathcal{F}}_{\rm rad}/h_\pm$. We note that it follows from Equations~(\ref{eq:magnetization}) to~(\ref{eq:skindepth}) that $\bar{\rho}^h= h_\pm\bar{\rho}$ and $\bar{d}^h=\bar{d}$. For $\tilde{\mathcal{F}}_{\rm rad}=\text{const.}$, and $h_\pm=\text{const.}$, the radiative transfer sources of Equations~(\ref{eq:momentumions}) to~(\ref{eq:photonmoments2}) can be neglected. As an illustrative example, we integrate the effective three-fluid system $\left(+,-,p\right)$ for boundary conditions of $\gamma_{u}=10$, with a magnetization of $\bar{\sigma}_{u}=10^{-6}$, and a multiplicity of $\mathcal{M}=1$. The discretized mesh spans $\tilde{x}\in\left[0,130\right]$ with $\Delta\tilde{x}=0.01$ and a time step according to the CFL condition of $f_{\rm CFL}=0.1$. Guided by the previously stated scalings with $h_\pm$, we vary $h_\pm$ itself and adapt the parameters $\chi$, $\mu$, and $\tilde{\mathcal{F}}_{\rm rad}$.  Figure~\ref{fig:ComparisonH} shows a direct comparison between the multi-fluid integration for cold plasma (top panel) and warm plasma (bottom panel) with parameters varied according to the aforementioned re-scaling. Especially, we see that for $h_\pm=\text{const.}$, we can adjust the linear-theory characteristics for the inter-species coupling in a heated plasma as follows:
\begin{align}
    \bar{\omega}_\pm^h&=h_\pm^{-1}\bar{\omega}_\pm\Rightarrow \lambda_\pm^h=h_\pm\lambda_\pm\label{eq:rescalingwarm}\,,\\
    \mathcal{A}_\pm^h&=\mathcal{A}_\pm\,.
\end{align}
While the amplitude of the lepton oscillation does not change, the length scale of the lepton coupling scales with the enthalpy. The following section will present a comparison of all relevant scales in self-consistently modeled deceleration profiles. Subsequent comparisons with the linear-theory estimates make use of the scaling in Equation~(\ref{eq:rescalingwarm}).

\subsection{Pair-loaded deceleration profiles}
\label{sec:pairdeceleration}

\begin{figure}
  \centering
  \includegraphics[width=0.475\textwidth]{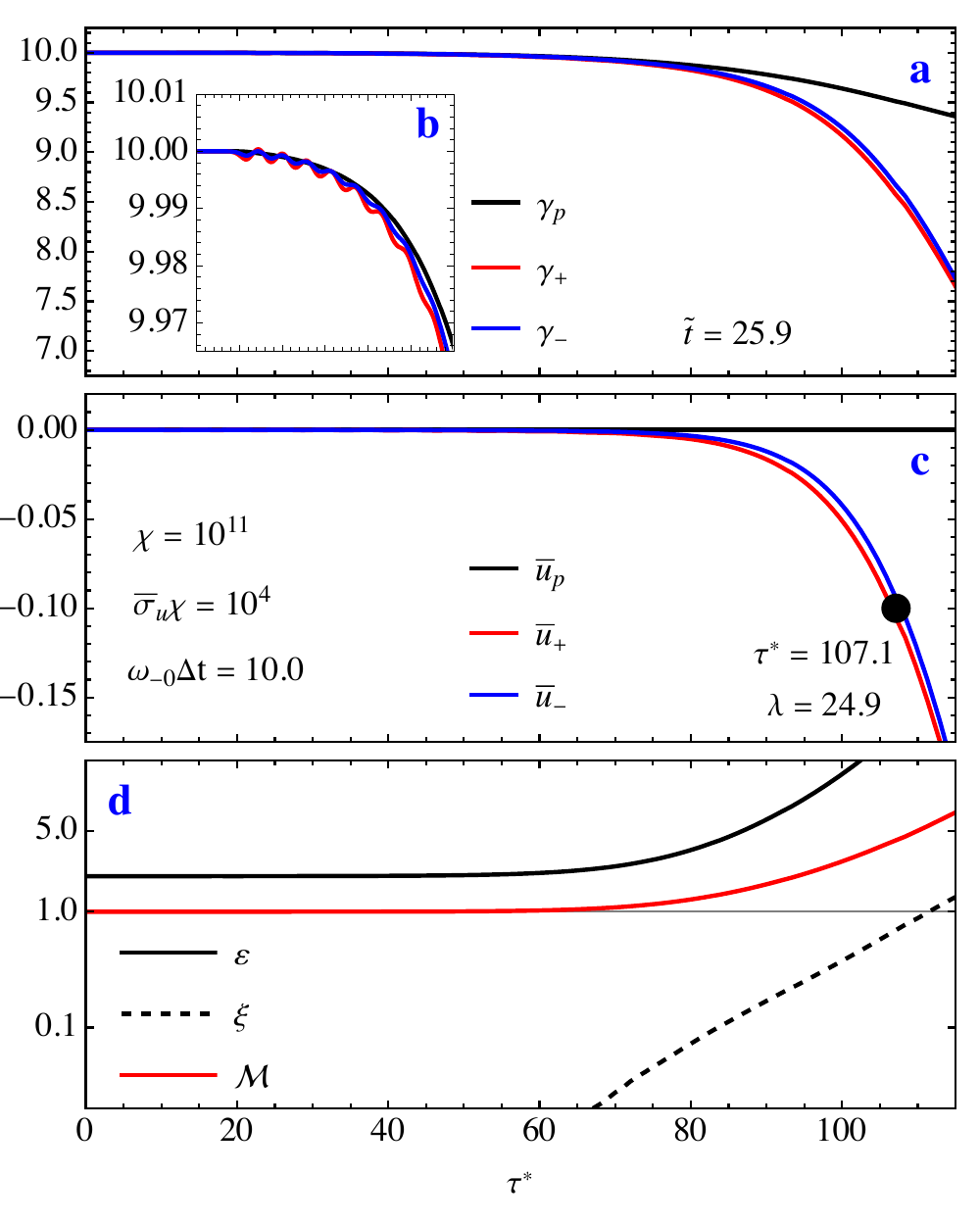}
  \vspace{-11pt}
  \caption{Solution of the magnetized multi-fluid integration (implicit-explicit) for \emph{weakly coupled} conditions with $\chi=10^{11}$,  $\bar{\sigma}_{u}=10^{-7}$ and $\omega_{-0}\Delta t=10.0$ at the time $\Bar{t}=25.9$. We display Lorentz factors of ions, electrons, and positrons (panel a) as well as a zoom into the $\tau^*\in\left[0,60\right]$ interval (panel b). For the chosen parameters, the deceleration length is comparable to the coupling wavelength, or $\lambda_R\lesssim\lambda_\pm$. Leptons and ions decouple and develop a velocity separation (panel c). We show the multiplicity profile as well as the scale ratios (panel d) defined in Equations~(\ref{eq:xicriterion}) and~(\ref{eq:decouplingcriterion}).}
\label{fig:solutionmagnetized}
\end{figure}

\begin{figure}
  \centering
  \includegraphics[width=0.475\textwidth]{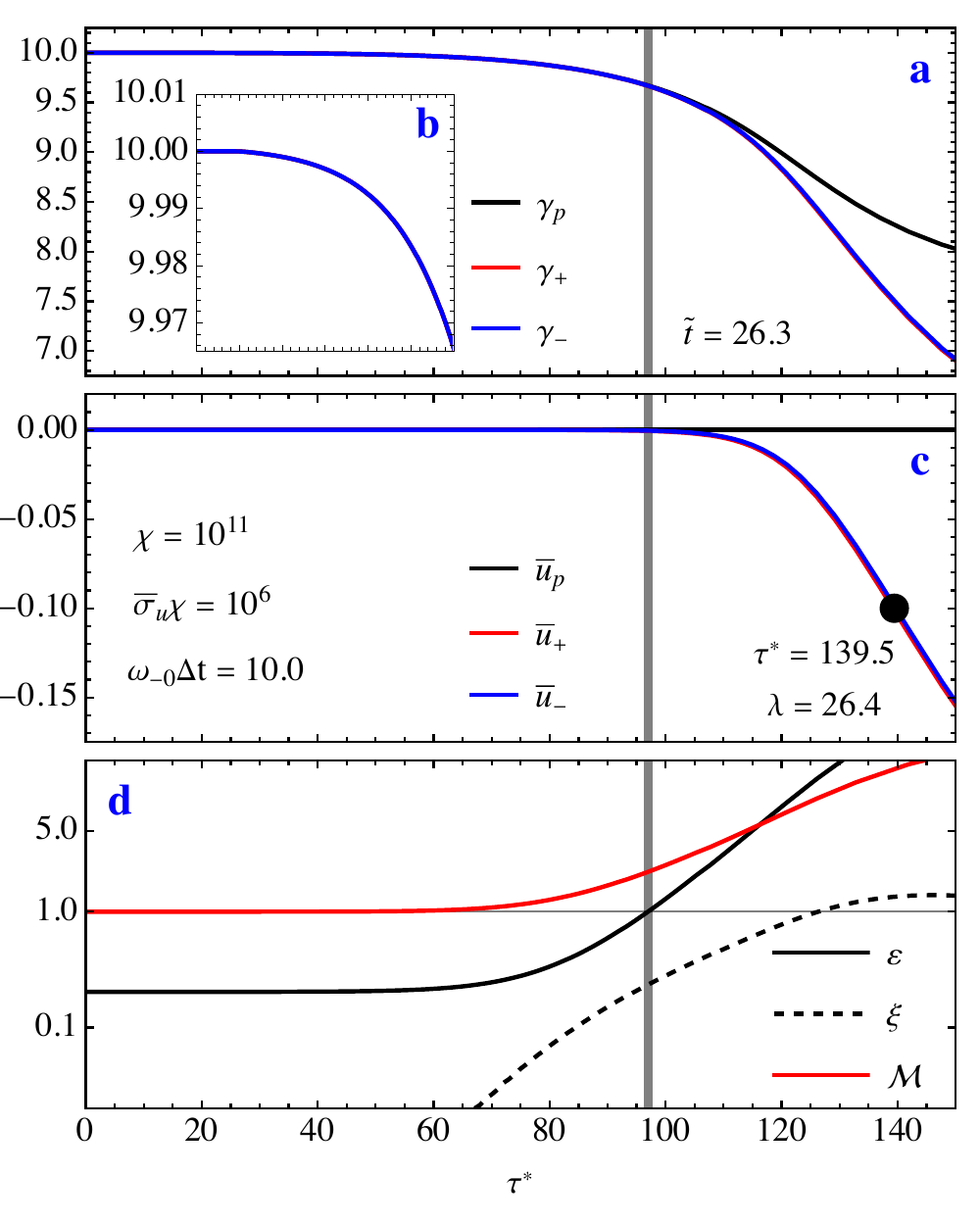}
  \vspace{-11pt}
  \caption{As Figure~\ref{fig:solutionmagnetized}, but for \emph{strongly coupled} conditions with a larger magnetization of $\bar{\sigma}_{u}=10^{-6}$ at the time $\Bar{t}=26.3$. For the chosen parameters, coupling initially dominates over deceleration, or $\lambda_R> \lambda_\pm$. The vertical grid line indicates the transition $\varepsilon=\;\lambda_\pm/\lambda_R=1$. The inset panel (b) extends over the interval $\tau^*\in\left[0,60\right]$.}
\label{fig:solutionmagnetized2}
\end{figure}

\begin{figure}
  \centering  \includegraphics[width=0.475\textwidth]{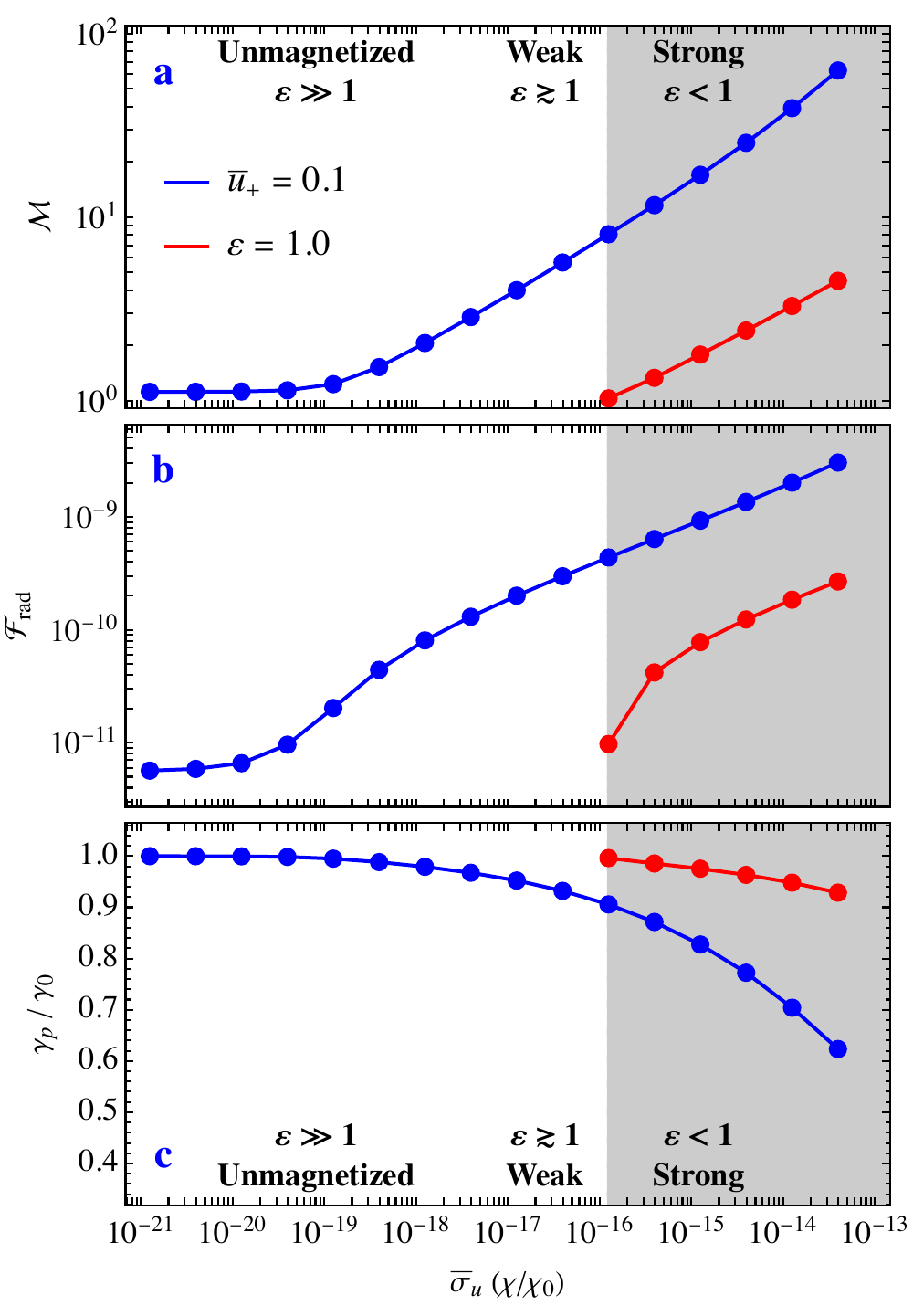}
  \vspace{-6pt}
  \caption{Multi-fluid integration (implicit-explicit) with varying magnetization for $\chi=10^{11}$ and $\omega_{-0}\Delta t=10.0$. We extract characteristic properties of the deceleration profiles at two different locations: for a fixed velocity separation of $\bar{u}_+=0.1$, and the location of decoupling, $\varepsilon=\;\lambda_\pm/\lambda_R=1$. This figure displays the multiplicity (panel a), the radiative force (panel b), as well as the baryon deceleration (panel c).}
\label{fig:magnetizedscan}
\end{figure}

\begin{figure}
  \centering
  \includegraphics[width=0.475\textwidth]{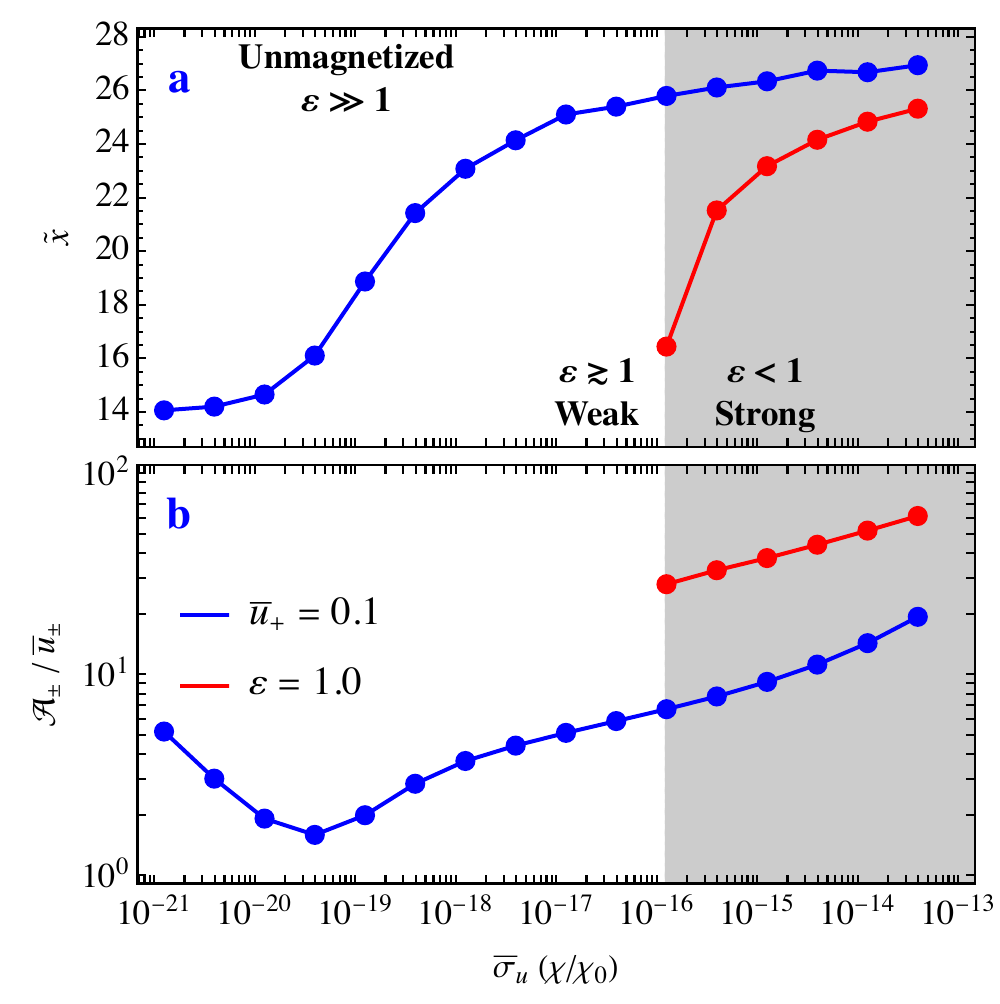}
  \vspace{-11pt}
  \caption{Selected properties of the deceleration profiles shown in Figure~\ref{fig:magnetizedscan}. Panel (a) shows the two positions where the measurements are taken and how they approach each other for increasing magnetization. We then compare the measured velocity separation to the linear-theory amplitude estimate of Equation~(\ref{eq:couplingamplitude}) in panel (b). Evaluating the expression for $\mathcal{A}_\pm$ overestimates the velocity separation throughout all tests, with the relative error increasing for higher magnetizations.}
\label{fig:magnetizedamp}
\end{figure}

In this section, we use $\tilde{\mathcal{F}}_{\rm rad}=2\tilde{\sigma}_\pm h_\pm u^x_\pm \tilde{n}_{\gamma\shortto u}$ as a radiative force profile in Equation~(\ref{eq:upmmomentum}). Such a choice of sources models the radiative transfer in the deceleration region self-consistently while capturing the effects of inter-species coupling by the transverse magnetic field. We integrate the full five-fluid system for boundary conditions of $\gamma_{u}=10$, a multiplicity of $\mathcal{M}=1$, a scale-separation of $\chi=10^{11}$, and a realistic mass ratio of $1/\mu=1836$. Following \citet{levinson2020b}, we seed the radiation field with $\tilde{n}_{\gamma\shortto u}(0)=0.01$. In this section, we vary the magnetization parameter $\bar{\sigma}_{u}$. The discretized mesh spans $\tilde{x}\in\left[0,40\right]$ with $\Delta\tilde{x}=0.005$ and uses a time step according to the CFL condition of $f_{\rm CFL}=0.1$.

To analyze the self-consistently modeled deceleration profiles, we introduce two relevant parameters for the comparison of various scales. First, we measure how rapidly the coupling length (Equation~\ref{eq:couplinglength}) changes per oscillation period of the inter-species coupling:
\begin{align}
\xi\equiv\frac{\text{d}\left(\lambda_\pm/\lambda\right)}{\text{d}\tilde{x}}\frac{1}{\left(\lambda_\pm/\lambda\right)}.
\label{eq:xicriterion}
\end{align}
Second, we compare the coupling length to the deceleration length scale. 
From the evaluation of unmagnetized deceleration profiles in the single-fluid approximation \citep{granot2018} we can deduce the deceleration length as
\begin{align}
    \lambda_R^{-1}=\left(1+4\eta\frac{\gamma^2}{\gamma_{u}}\right)\tilde{\sigma}_{\gamma\gamma} \left(\tilde{n}_{l}+1\right)\frac{1}{\lambda}\;.
\end{align}
We note that in the deceleration region, before the decoupling of species, the different cross sections $\tilde{\sigma}_\pm\approx\tilde{\sigma}_{\gamma\gamma}$ coincide with good accuracy. We can then compare the coupling length scale to the local deceleration scale and for $\tilde{n}_{l}\approx3\mathcal{M}\gg1$ we find
\begin{align}
    \varepsilon=\frac{\lambda_\pm}{\lambda_R}\approx12\pi\frac{\gamma^2}{\gamma_{u}}\left(\frac{\Tilde{B}_{u}}{\Tilde{B}_z}\right)\mathcal{M}^2\left(1+4\eta\frac{\gamma^2}{\gamma_{u}}\right)h_\pm\tilde{\sigma}_{\gamma\gamma}\frac{\bar{\rho}}{\lambda}.
    \label{eq:decouplingcriterion}
\end{align}
The derivation of expression (\ref{eq:decouplingcriterion}) uses the scaling with enthalpy established in Section~\ref{sec:couplingomega}. It is convenient to use $\varepsilon$ for the definition of different coupling regimes. In the unmagnetized case, radiative deceleration dominates all species. This limit corresponds to $\varepsilon\gg 1$. Weak inter-species coupling is expected when $\lambda_\pm$ approaches the radiative length scale, namely $\varepsilon\gtrsim 1$. Finally, we define strong coupling as the limit where the deceleration length exceeds the typical scale for lepton-ion coupling, as to say $\varepsilon<1$.

Figure~\ref{fig:solutionmagnetized} shows a typical deceleration profile for a mild magnetization $\bar{\sigma}_{u}\chi=10^4$ and a weak inter-species coupling. As the coupled lepton-ion fluid decelerates, the multiplicity gradually grows (panel d). Eventually, the species decouple and a significant velocity separation is established (panel c). The lepton-ion coupling can be identified by the lepton oscillations in panel (b) of Figure~\ref{fig:solutionmagnetized}. As expected from Equations~(\ref{eq:couplinglength}) and~(\ref{eq:couplingamplitude}) both the coupling amplitude and wavelength are expected to decrease for larger magnetization. Therefore, the lepton oscillations are not directly noticeable for the strongly coupled case displayed in Figure~\ref{fig:solutionmagnetized2}. Comparing both setups, we outline the main effects of magnetic coupling. First and foremost, the coupling of species due to higher magnetization shifts the location of significant velocity separation downstream. For the presented cases, the location of a velocity separation of $\bar{u}_+=0.1$ is translated by $\Delta\lambda\approx 1.4$ and $\Delta\tau^*=32.3$. Following the increasing optical depth, the multiplicity at this given velocity separation changes by $\Delta\mathcal{M}\approx 13$. At the same time, the radiative force increases by a factor of $4.5$. 

Figure~\ref{fig:magnetizedscan} repeats this analysis more systematically for a wide scan of magnetization parameters, spanning over seven orders of magnitude. By varying the upstream magnetization we effectively change the upstream value of $\varepsilon$, as to say the strength of coupling prior to the deceleration. When initializing $\varepsilon<1$ (similar to the case shown in Figure~\ref{fig:solutionmagnetized2}), we can use the transition point $\varepsilon=1$ to measure the location of lepton-ion decoupling. We indicate the transition between the coupling regimes (see Equation~\ref{eq:decouplingcriterion} and below) by a gray shade in Figure~\ref{fig:magnetizedscan}. As expected from the linear estimate in Equation~(\ref{eq:couplingamplitude}), an increasing magnetization requires a larger multiplicity and radiative force to reach a certain velocity separation. This statement is true independently of the measurement location in the immediate deceleration region, with similar trends at $\bar{u}_+=0.1$ (blue line) and $\varepsilon=1$ (red line). However, as we show in Figure~\ref{fig:magnetizedamp} (panel b), the amplitude estimate of Equation~(\ref{eq:couplingamplitude}) broadly overestimates the actual velocity separation. The relative difference between a linear-theory estimate \citep[see also][]{Vanthieghem2022} and the actual separations in self-consistent deceleration profiles grows with increasing magnetization. 

The principal reason for the differences emerging in Figure~\ref{fig:magnetizedamp} can be found in Equation~(\ref{eq:xicriterion}). The linear analysis reproduced in Appendix~\ref{app:linearanalysis} relies on the assumption that the wavelength of the lepton-ion coupling changes slowly. In other words, for Equation~(\ref{eq:couplingamplitude}) to be applicable requires $\xi\ll 1$. Already for mild magnetizations, this criterion is not fulfilled in the deceleration region (dashed line in panel d, Figures~\ref{fig:solutionmagnetized} and~\ref{fig:solutionmagnetized2}). With increasing magnetization, we generally find $\xi\gtrsim 1$. This finding has significant consequences. While lepton oscillations can be easily found for mild magnetization and small multiplicity (panel b, Figure~\ref{fig:solutionmagnetized}), the rapid change of $\lambda_\pm$ renders the linear theory reviewed in Section~\ref{sec:couplingomega} inaccurate during the strongest deceleration. This especially applies to the estimate of the oscillation amplitude in Equation~(\ref{eq:couplingamplitude}).

Leptons and ions decouple when the coupling length exceeds the deceleration length, independently of the amplitude of lepton oscillations. As argued above, the role of the magnetic field becomes subdominant for $\varepsilon>1$. Thus, a (sufficient) condition for decoupling can be obtained by $\varepsilon=1$. Figure~\ref{fig:solutionmagnetized2} shows the coincidence of $\varepsilon=1$ with the onset of the rapid growth of multiplicity (panel d) and (eventually) velocity separation (panel c). We, therefore, use $\varepsilon=1$ as a second criterion to evaluate plasma properties during the decoupling phase in Figure~\ref{fig:magnetizedscan}. There is a significant separation between all observables at the different measurement points. At the same time, as shown in Figure~\ref{fig:magnetizedamp} (panel a), the location of decoupling ($\varepsilon = 1$) approaches the one where $\bar{u}_\pm=0.1$. During the most efficient deceleration, the density of quanta grows rapidly and all species slow down due to the action of the radiative drag force. Therefore, the possible length of the deceleration region is finite, and we discuss the implications of these findings in the following section. 

\section{Discussion}
\label{sec:discussion}

\subsection{Thresholds for magnetic coupling in the deceleration region}
\label{sec:thresholds}

\begin{figure}
  \centering
  \includegraphics[width=0.475\textwidth]{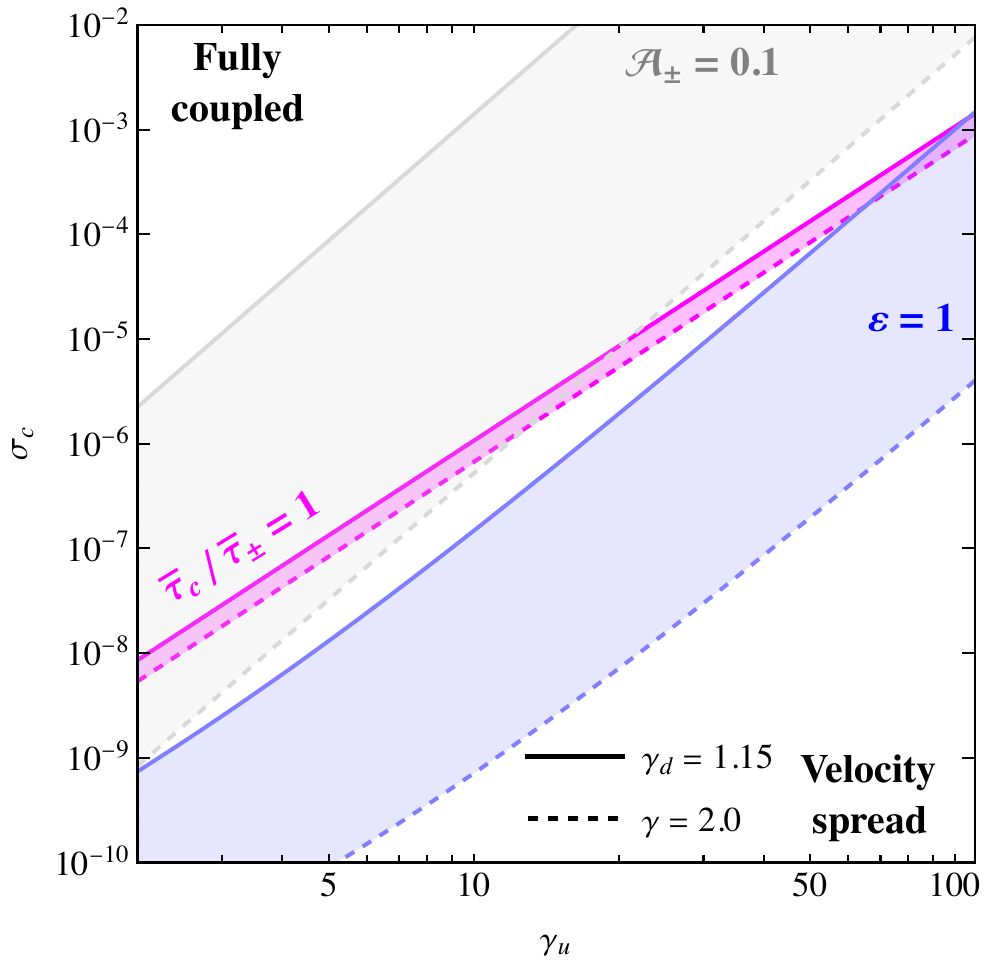}
  \vspace{-11pt}
  \caption{Different estimates of the critical magnetization for varying injection velocities $\gamma_{u}$. We visualize, which combinations of $\gamma_{u}$ and $\sigma_{c}$ potentially lead to sufficient velocity separation and the development of microturbulence \citep{Vanthieghem2022}. The conservative limit given by the turbulent coupling $\bar{\tau}_{c}/\bar{\tau}_\pm=1$ in Equation~(\ref{eq:decouplingmodel}) is displayed by blue lines. Conservative limits given by the growth rate of the instability (Equation~\ref{eq:sigmacgrate}) are shown by magenta lines. Gray lines denote the previously employed but invalid linear theory estimate (Equation~\ref{eq:couplingamplitude}).}
\label{fig:magnetizedlimits}
\end{figure}

In Section~\ref{sec:pairdeceleration} we outline various indicators that show the insufficiency of the coupling amplitude $\mathcal{A}_\pm$ (Equation~\ref{eq:couplingamplitude}) as a measure of velocity separation. In general, $\mathcal{A}_\pm$ overestimates the velocity separation by more than one order of magnitude. In other words, the self-consistent deceleration profiles create a velocity separation $\bar{u}_\pm$ for combinations of multiplicity $\mathcal{M}$ and radiative force $\mathcal{F}_{\rm rad}$ incompatible with Equation~(\ref{eq:couplingamplitude}). A closer analysis of the separation of scales shows that during the deceleration phase, we find $\xi\approx 1$. This regime of rapidly changing coupling properties naturally renders the amplitude estimate $\mathcal{A}_\pm$ inaccurate, and we require an adapted limit to the critical magnetization.

We use Equation~(\ref{eq:decouplingcriterion}) to constrain a critical magnetization $\sigma_{c}$ that discriminates between coupled profiles and those with significant velocity separation. A conservative estimate can be developed by assuming that the multiplicity saturates at maximum downstream values of $\mathcal{M}\approx \gamma_{u}/(3\mu)\gg 1$ \citep{ito2020}, that the magnetic field remains $\Tilde{B}_{u}\approx \Tilde{B}_z$, and that $\tilde{n}_{p}\approx 1$. For species decoupling at $\varepsilon \approx 1$, Equation~(\ref{eq:decouplingcriterion}) can be written as
\begin{align}
    &\sigma_{c}\left(\gamma_d\right)\approx 9.7\times 10^{-23} \gamma_d^4 \mathcal{M}^4 h_\pm^2\tilde{\sigma}_{\gamma\gamma}\left(1+4\eta\frac{\gamma_d^2}{\gamma_{u}}\right)\,.
\label{eq:decouplingmodel}
\end{align}
For a decoupling velocity of $\gamma_d=1.15$ and $\gamma_{u}\gg 1$ we find
\begin{align}
    \sigma_{c}\left(\gamma_d=1.15,\gamma_{u}\gg 1\right)\approx 10^{-7}\left(\frac{\gamma_{u}}{10}\right)^4\,.
\end{align}
\citet{Vanthieghem2022} estimate the turbulence coupling time required for the microturbulence to grow up to a level that can balance the radiation drag as $\bar{\omega}_{pu}\bar{\tau}_{c}=10\sqrt{\mathcal{M}h_\pm \mu/\mathcal{F}_{\rm rad}}$. We obtain the time scale $\bar{\omega}_{pu}\bar{\tau}_\pm$ for the magnetic coupling from Equation~(\ref{eq:couplinglength}), and requiring $\bar{\tau}_{c}/\bar{\tau}_\pm=1$ yields a separate threshold for $\sigma_{c}$:
\begin{align}
    &\sigma_{c}\left(\gamma_d\right)\approx 4.3\times 10^{-13}\gamma_{u}^2\gamma_d^2\mathcal{M}h_\pm\,.
    \label{eq:sigmacgrate}
\end{align}
For a decoupling velocity of $\gamma_d=1.15$ and $\gamma_{u}\gg 1$ we find
\begin{align}
    \sigma_{c}\left(\gamma_d=1.15,\gamma_{u}\gg 1\right)\approx 10^{-6}\left(\frac{\gamma_{u}}{10}\right)^3\,.
\end{align}
The critical magnetization constrained by Equation~(\ref{eq:decouplingmodel}) assumes that the decoupling happens close to the immediate downstream. However, merely reaching $\varepsilon=1$ does not imply a sufficient velocity separation of the different species to trigger plasma instabilities and microturbulence. The system needs time to develop a significant velocity spread in the decoupled state. Specifically, when $\varepsilon\gg 1$, the action of the radiative force becomes dominant and the system effectively behaves like the unmagnetized one. The growth of the velocity separation is then proportional to the radiative force $\mathcal{F}_{\rm rad}$ and develops on a fraction of the radiative length scale. While the exact velocity during the decoupling phase is unknown, we repeat our previous analysis for conditions where the effective one-fluid system has not fully decelerated. \citet{granot2018} estimate the number of quanta for a mildly relativistic RMS profile as
\begin{align}
    \mathcal{M}\approx\frac{\tilde{n}_{l}}{3}\approx\frac{1}{12\mu\eta}\frac{\gamma_{u}}{\gamma^2}\,.
\end{align}
For an arbitrarily chosen value of $\gamma=2$ to allow for a velocity separation before full deceleration of all species, Equation~(\ref{eq:decouplingmodel}) becomes
\begin{align}
    \sigma_{c}\left(\gamma=2\right)&\approx 1.9\times 10^{-13}\gamma_{u}^4 \,.
    \label{eq:sigmagammatwo}
\end{align}
The thresholds for decoupling at different locations in the deceleration profiles derived in Equations~(\ref{eq:decouplingmodel}) and~(\ref{eq:sigmagammatwo}) are highly sensitive to the velocity during decoupling (with a dependence on the fourth power on the flow Lorentz factor). They differ by more than two orders of magnitude. Figure~\ref{fig:magnetizedlimits} gives a visual account of the various estimates of $\sigma_{c}$ presented in this section. The shaded region enclosed by blue lines is the main novelty of this paper, fixing $\sigma_{c}$ by evaluating when the multi-species beam decouples for $\gamma\lesssim 2$ at $\varepsilon=1$. Contrasting this conservative estimate that is validated by the self-consistent deceleration profiles presented in Section~\ref{sec:pairdeceleration}, the linear theory threshold derived from the coupling amplitude (\ref{eq:couplingamplitude}) over-estimates $\sigma_{c}$ by several orders of magnitude (gray shaded region). Finally, magnetization thresholds from demanding sufficiently fast growth of the seed instabilities of the microturbulence discussed in \citet[][see Figure~2]{Vanthieghem2022} are indicated by magenta-colored bands.

\subsection{Magnetization of astrophysical environments}

Relativistic RMS may take place in different astrophysical environments with very different magnetization levels. One such environment is the ejecta from binary neutron star or black hole-neutron star mergers, where the shock driven by the GRB jet and the cocoon is mildly relativistic and possibly ultra-relativistic. The magnetic field in the ejecta is unknown, but we can get a rough estimated lower limit as follows. The expectation is that at the launching site of the ejecta, at radii $r \sim 10^6-10^7$ cm,  the magnetic field is not very far from equipartition. As the ejecta expand the density of each fluid element drops as $r^{-3}$ while the magnetic field energy density drops at most as $r^{-4}$. This implies that $\sigma$ drops at most as $r^{-1}$, so we can estimate $\sigma \gtrsim (10^6 {\rm ~cm})/r$. Therefore, any shock that takes place at $r \lesssim 10^{13}$ cm (i.e., within $\sim 1000$ s of the merger) will have a magnetic field that is strong enough for full coupling.

Relativistic shocks can also occur in extreme supernovae of stripped-envelope progenitors. Such shocks can be driven by a highly energetic explosion or by a jet and its cocoon, such as a long GRB jet  \citep{nakar2012}. These shocks can propagate either within the stellar envelope or in the wind. The magnetization level is 
\begin{align}
\sigma \sim 10^{-8} \left(\frac{B}{100~G}\right )^2 \left(\frac{\rho}{10^{-10}{\rm g\,cm^{-3}} }\right )^{-1} \gamma_u^{-2}\;.
\end{align}
The magnetic field in these progenitors is unknown, but we can use observations of Wolf-Rayet stars to get an idea of possible values. These observations suggest that the magnetic field on the surface can be as high as a few hundred Gauss \citep[e.g.,][]{Chevrotiere2014,Hubrig2016}. The upstream density depends on whether the shock travels in the envelope or the wind. The density in the envelope drops near its edge and at an optical depth of unity (to infinity) it is of the order of $10^{-7} {\rm g\,cm^{-3}}$ for a star with a mass of $10M_\odot$ and radius of $10^{11}$ cm \citep[e.g.,][]{Nakar2010}. Therefore, a relativistic shock that propagates near the edge of the envelope of a compact Wolf-Rayet star probably does not have full coupling between the various plasma constituents. In the wind, the density is expected to be of the order of $10^{-10} {\rm g\,cm^{-3}}$ near the stellar edge at a radius of $\sim 10^{11}$ cm, implying that the magnetic fields are more likely to induce full coupling. Finally, in low-luminosity GRBs, the shock seems to break out of the progenitor at a radius of $\sim 10^{13}$ cm \citep{Nakar2015}. In these progenitors, there is no information about the magnetic fields. 

\subsection{Limitations}

In this paper, we study the coupling of electron-positron pairs to a single species of heavy ions during the formation of relativistic radiation-mediated shocks in the presence of a magnetic field perpendicular to the initial propagation direction of the plasma flow. In particular, we provide critical magnetizations $\sigma_{c}$ that effectively prevent the radiation drag from decoupling pairs and ions for different upstream velocities $\gamma_{u}$ (Figure~\ref{fig:magnetizedlimits}). Relativistic shocks ($\gamma_{u}\gg 1$) have magnetic fields oriented quasi-perpendicular to the flow direction expansion \citep[][see also discussion in \citealt{Plotnikov_2018}]{Gallant1992}, for which the 1D models shown in this work are a viable approximation. Mildly relativistic flows can show a significant longitudinal magnetic field component with $B_\parallel$ along $\mathcal{F}_{\rm rad}$. Parallel magnetic fields do not affect the longitudinal fluid velocities of different species, and as such, they do not contribute to the coupling of different species examined in this paper. However, the additional transverse currents they induce can change the nature and growth of the micro-instabilities described by \citet{Vanthieghem2022}. How the radiation-drag-induced microturbulence reacts to magnetic fields for various obliquities can be treated as an isolated question in future work. We predict the excitation of microturbulence due to plasma instabilities for low magnetization. In this regime, the effect of transverse magnetic fields on the growth of the microinstabilities is negligible. In the regime of strong coupling, transverse instabilities can develop in addition to magnetically driven oscillations. The growth of such modes cannot be captured in our reduced 1D description and poses a clear limitation. However, their saturation level is likely constrained by the absolute velocity separation between species for a given external magnetic field. In this work, we find that the velocity separation is small for large parts of the deceleration region and only becomes relevant close to the immediate downstream. The critical magnetization levels derived in Section~\ref{sec:thresholds} are conservative estimates of the conditions during decoupling that likely occur for small velocity separations (see Figure~\ref{fig:solutionmagnetized2}). Studying the feedback of additional transverse modes on the shock dynamics is left for further studies with consistent shock (micro)physics and higher dimensionality.

Our derivations \citep[and those by][]{Vanthieghem2022} consider the dynamics of a baryon-pair plasma with a single ion species. As visualized in Figure~\ref{fig:shematics} (panels d/e), such a three-component plasma efficiently re-accelerates pairs in the presence of transverse magnetic fields and counteracts the radiation drag. Such a well-coupled three-component plasma can be treated in a single fluid approximation and will likely suppress kinetic instabilities due to negligible velocity spreads. This simple picture becomes more complicated when ion species with different charge-to-mass ratios are present. Mediated by the lepton deceleration, the various ion species will quickly develop a velocity separation, and pairs will no longer be coupled to the single stream of heavy ions. Velocity separations will likely develop even in the presence of transverse magnetic fields. However, the dynamics of these magnetized multi-ion systems as well as their unstable modes are not yet explored.

\section{Conclusion}

The main result of this paper provides critical magnetizations for which the growth of kinetic instabilities in a photon-rich three-species pair-ion plasma becomes increasingly difficult. Consistently derived limits on $\sigma_{c}$ (blue band in Figure~\ref{fig:magnetizedlimits}) are lower than our previous simplistic estimates \citep[][Section 5]{Vanthieghem2022}. In the presence of a transversal magnetic field, especially for relativistic systems with upstream velocities of $\gamma_{u}\approx 10$, a mild magnetization of $\sigma_{c}\approx 10^{-7}$ is sufficient to prevent velocity spreads and the growth of kinetic instabilities. The magnetization in different RMS environments is not sufficiently constrained by theory or observation. Still, the presence of at least small magnetic fields is very likely, such that plasma components may decouple only in the late phase of their deceleration, close to the subshock in the downstream transition. There, the growth of microturbulence would induce high-energy tails in the particle distribution with a possible imprint on the signal observed during the shock breakout. While we do not know what an imprint of microturbulence in the shock breakout signal would look like, its occurrence, or lack thereof, can now be used to constrain the ambient magnetic field strength.

\section{Data Availability}
The data underlying this article will be shared on reasonable request to the corresponding author.

\section{Acknowledgments}
This research is supported in part by NASA grant 80NSSC18K1099 and NSF grant PHY-2206607. AAP and JFM acknowledge support from the National Science Foundation (NSF) under grant AST-1909458. This research was facilitated by Multimessenger Plasma Physics Center (MPPC), NSF grant PHY-2206607. AV acknowledges support from the NSF grant AST-1814708 and the NIFS Collaboration Research Program (NIFS22KIST020). FF acknowledges support by the US DOE FES under FWP-100742. AL and EN acknowledge support from the Israel Science Foundation grant 1995/21. The presented numerical simulations were conducted on the \emph{stellar} cluster (Princeton Research Computing). AV and JFM thank Tel Aviv University for their generous hospitality. 


\bibliographystyle{mnras}
\bibliography{literature} 



\appendix

\section{Linear analysis of the multi-species plasma response}
\label{app:linearanalysis}

\begin{figure}
  \centering
  \includegraphics[width=0.475\textwidth]{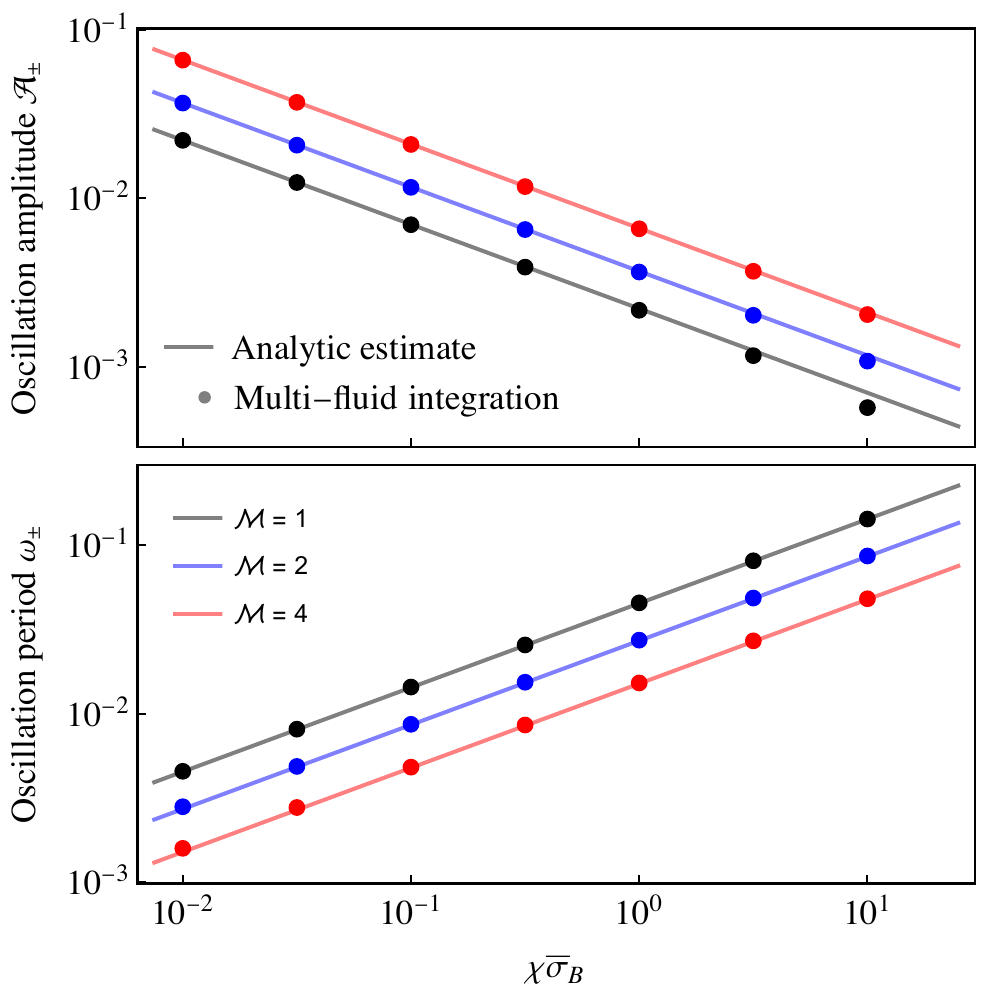}
  \vspace{-6pt}
  \caption{Validation of the linear analysis for the amplitude $\mathcal{A}_\pm$ of the velocity spread (top panel) and coupling frequency $\omega_\pm$ (bottom panel). For different multiplicities $\mathcal{M}$, we show measurements from a direct (explicit) integration of the multi-fluid system (data points) for $\gamma_{u}=10$,  $\bar{\omega}_{p0}=0.1$ and $\chi=10^7$. The data points are compared to the analytic estimates (solid lines) given by Equations~(\ref{eq:couplinglength}) and~(\ref{eq:couplingamplitude}).}
\label{fig:Validation}
\end{figure}

This section reviews the linear analysis of the characteristic response of a magnetized multi-species plasma vastly used in this work and parts of \citet[][]{Vanthieghem2022}. As shown below, the allowed plasma oscillations are independent of the radiative force $\tilde{\mathcal{F}}_{\rm rad}$, assumed to be constant in this section. We further assume that the system is uniform ($k_x=0$), that the plasma is cold ($h_\pm=1$), and that the mass ratio is infinite, $\mu\rightarrow 0$. Without loss of generality, we can reduce the system of Equations~(\ref{eq:momentumions}) to~(\ref{eq:couplingIII}) and assume that all quantities are given in the plasma rest frame:
\begin{align}
    \partial_{\tilde{t}} u^x_s &=\frac{\tilde{q}_s}{\tilde{m}_s}\left(\tilde{E}_x+\beta^y_s \tilde{B}_z\right)-\delta_s\tilde{\mathcal{F}}_{\rm rad} \,,\label{eq:reducedI}\\
    \partial_{\tilde{t}} u^y_s&=\frac{\tilde{q}_s}{\tilde{m}_s}\left(\tilde{E}_y-\beta^x_s \tilde{B}_z\right) \,,
\\
    \partial_{\tilde{t}}\tilde{E}_x&=-\chi\sum_s \tilde{q}_s \bar{n}_s \beta^x_s \,,\\
    \partial_{\tilde{t}}\tilde{E}_y&=-\chi\sum_s \tilde{q}_s \bar{n}_s \beta^y_s\label{eq:reducedIV}\,.
\end{align}
Here, we used the normalization $\tilde{q}_s=q_s/e$ and $\tilde{m}_s=m_s/m_e$ for the different species $s$. We define $\delta_{p}=0$ and $\delta_\pm=1$ to study the allowed perturbations of the equilibrium flow of $\gamma_0$. We exploit a linearization of velocities $\mathbf{u}_s=\gamma_s \boldsymbol{\beta}_s$, namely $\delta u_s^x=\gamma_0^3\delta\beta_s^x$ and $\delta u_s^y=\gamma_0\delta\beta_s^y$. Since in the limit $\mu\rightarrow 0$ the change in the ion velocity vanishes, $\delta u_{p}^i=0$, only $s\in\left[+,-\right]$ is considered and in frequency space, where we use $\partial_{\tilde{t}}\rightarrow - i\omega$, Equations~(\ref{eq:reducedI}) to~(\ref{eq:reducedIV}) become
\begin{align}
    -i\bar{\omega} \delta u^x_s &=\frac{\tilde{q}_s}{\tilde{m}_s}\left(\delta \tilde{E}_x+\frac{\delta u^y_s}{\gamma_0^2} \frac{\bar{\lambda}}{\bar{\rho}}\right)-\tilde{\mathcal{F}}_{\rm rad} \,,\\
    -i\bar{\omega} \delta u^y_s&=\frac{\tilde{q}_s}{\tilde{m}_s}\left(\delta \tilde{E}_y-\frac{\delta u^x_s}{\gamma_0^4} \frac{\bar{\lambda}}{\bar{\rho}}\right) \,,
\\
    -i\bar{\omega} \delta\tilde{E}_x&=-\frac{\chi}{\gamma_0}\sum_s \tilde{q}_s \bar{n}_s \delta u^x_s \,,\\
    -i\bar{\omega}\delta\tilde{E}_y&=-\frac{\chi}{\gamma_0}\sum_s \tilde{q}_s \bar{n}_s \delta u^y_s\label{eq:transversecurrent}\,.
\end{align}
This inhomogeneous system of equations admits a particular solution for which all time derivatives vanish:
\begin{align}
    \delta u^y_- &= \frac{\bar{n}_-}{\bar{n}_+}\delta u^y_+=\frac{\mathcal{M}}{\mathcal{M}+1}\delta u^y_+ \,,\\
    \delta u^y_+ &= 2\gamma_0^2\left(\mathcal{M}+1\right)\frac{\bar{\rho}}{\bar{\lambda}}\tilde{\mathcal{F}}_{\rm rad} \,,\\
    \delta\tilde{E}_x &= -\left(2\mathcal{M}+1\right)\tilde{\mathcal{F}}_{\rm rad}\,.
\end{align}
Here we use $\delta u^x_s = \delta\tilde{E}_y = 0$, with the plasma multiplicity $\mathcal{M}=\bar{n}_+$. This particular solution dictates the amplitude of the oscillations of the general solutions (see below). The appearance of the constant longitudinal electric field $\delta\tilde{E}_x$ is imposed by requiring a vanishing mean transverse current (i.e., no constant current) in Equation~(\ref{eq:transversecurrent}). In reality, when the ion mass is finite, this mean electric field is responsible for the deceleration of the ions (and hence the entire plasma) inside the shock transition layer. The general solution is the sum of the particular and the homogeneous
solutions, where the homogeneous solution is
\begin{align}
    -i\bar{\omega} \delta u^x_s &=\frac{\tilde{q}_s}{\tilde{m}_s}\left(\delta \tilde{E}_x+\frac{\delta u^y_s}{\gamma_0^2} \frac{\bar{\lambda}}{\bar{\rho}}\right)\,,\label{eq:linearI}\\
    -i\bar{\omega} \delta u^y_s&=\frac{\tilde{q}_s}{\tilde{m}_s}\left(\delta \tilde{E}_y-\frac{\delta u^x_s}{\gamma_0^4} \frac{\bar{\lambda}}{\bar{\rho}}\right)\,,
\label{eq:linearII}\\
    -i\bar{\omega} \delta\tilde{E}_x&=-\frac{\chi}{\gamma_0}\sum_s \tilde{q}_s \bar{n}_s \delta u^x_s\,,\label{eq:linearIII}\\
    -i\bar{\omega}\delta\tilde{E}_y&=-\frac{\chi}{\gamma_0}\sum_s \tilde{q}_s \bar{n}_s \delta u^y_s\label{eq:linearIV}\,.
\end{align}
The linearized momentum Equations~(\ref{eq:linearI}) and~\ref{eq:linearII} can be cast in matrix form
\begin{align}
\begin{split}
    \begin{pmatrix}
    \delta\Tilde{E}_x \\
    \delta\Tilde{E}_y
    \end{pmatrix} &=
    \begin{pmatrix}
    -i\bar{\omega}\left(\tilde{m}_s/\tilde{q}_s\right) & -1/\gamma_0^2\times\bar{\lambda}/\bar{\rho} \\
     1/\gamma_0^4\times\bar{\lambda}/\bar{\rho} & -i\bar{\omega}\left(\tilde{m}_s/\tilde{q}_s\right)
    \end{pmatrix} 
    \begin{pmatrix}
    \delta u_s^x \\
    \delta u_s^y
    \end{pmatrix} \equiv
    \mathcal{M}_s
    \begin{pmatrix}
    \delta u_s^x \\
    \delta u_s^y
    \end{pmatrix}\,,
\end{split}
\end{align}
with the inverse
\begin{align}
    \mathcal{M}_s^{-1}=    \frac{\gamma_0}{\omega^2-\Omega_s^2}\frac{\tilde{q}_s}{\tilde{m}_s}\begin{pmatrix}
    i\omega & -\Omega_s\gamma_0 \\
     \Omega_s/\gamma_0 & i\omega
    \end{pmatrix}\,.
\end{align}
Here, we introduced the \emph{lab-frame} quantities $\omega=\gamma_0\bar{\omega}$ and $\gamma_0^2\Omega_s=\left(\tilde{q}_s/\tilde{m}_s\right)\bar{\lambda}/\bar{\rho}$. We can now write $\delta\mathbf{u}_s=\mathcal{M}_s^{-1}\cdot\delta\tilde{\mathbf{E}}$, such that Equations~(\ref{eq:linearIII}) and~(\ref{eq:linearIV}) become
\begin{align}
    -i{\omega}\delta\tilde{\mathbf{E}}=-\chi\sum_s\bar{n}_s\tilde{q}_s\delta\mathbf{u}_s=-\chi\sum_s\bar{n}_s\tilde{q}_s\mathcal{M}_s^{-1}\cdot\delta\tilde{\mathbf{E}}\,.
\end{align}
Rearranging this expression yields
\begin{align}
    \left[\mathbb{1}+i\sum_s\frac{\chi\bar{n}_s\tilde{q}_s}{\omega}\mathcal{M}_s^{-1}\right]\cdot\delta\Tilde{\mathbf{E}}\equiv\xi\cdot\delta\Tilde{\mathbf{E}} =0.
\end{align}
Non-trivial solutions to this equation require $\text{det}\;\xi=0$. We exploit a simpler representation of $\xi$, namely
\begin{align}
    \xi=\begin{pmatrix}
    S & -iD\gamma_0\\
    iD/\gamma_0 & S
    \end{pmatrix}\qquad \text{det}\;\xi=S^2-D^2\,.
\end{align}
Here, we use ${\omega}_{s}^2=\chi\gamma_0 \tilde{q}_s^2 \bar{n}_s/\tilde{m}_s$ to express the variables $S$ and $D$:
\begin{align}
        S&= 1-\sum_s\frac{{\omega}_{s}^2}{\omega^2-\Omega_s^2}\,,\\
    D&= \sum_s\frac{\Omega_s}{\omega}\frac{{\omega}_{s}^2}{\omega^2-\Omega_s^2}\,.
\end{align}
We approximate this to first order by Taylor expansion around $\Omega_s / \omega\approx 0$. Hence, we consider the regime in which the frequency of the instability exceeds the gyration frequency of all species.
\begin{align}
    S& \approx 1-\sum_s\frac{\omega_{s}^2}{\omega^2}\,,\label{eq:Sapprox1}\\
    D& \approx \sum_s\frac{\omega_{s}^2}{\omega^2}\frac{\Omega_s}{\omega}\,.\label{eq:Sapprox2}
\end{align}
From $\text{det}\;\xi=(S-D)(S+D)$ it is straightforward to find $D=\pm S$. This equation has three roots for the system defined in Equations~(\ref{eq:Sapprox1}) and~(\ref{eq:Sapprox2}). First, we explore the limit of $\omega\approx\omega_0$. At zeroth order in $\Omega_s / \omega\approx 0$ one has $D_0\approx 0$, such that
\begin{align}
    S_0=1-\sum_s\frac{\omega_{s}^2}{\omega_0^2}\approx 0\,.\label{eq:Szeroth}
\end{align}
At this point, we recall that we assume a plasma of multiplicity $\mathcal{M}=\tilde{n}_+$, and $\tilde{n}_{-}=\mathcal{M}+1$. Equation~(\ref{eq:Szeroth}), thus, reduces to the pair plasma frequency
\begin{align}
    \omega_0^2=\sum_s\omega_{s}^2=\chi\gamma_0\left(1+2\mathcal{M}\right)\label{eq:Omega0}\,.
\end{align}
In the same way, we can generalize the first-order expression in Equation~(\ref{eq:Sapprox2}):
\begin{align}
    D\approx -\frac{1}{\gamma_0}\frac{\chi}{\omega^3}\frac{\bar{\lambda}}{\bar{\rho}}\,.
\end{align}
We now aim at obtaining an expression for $\omega$ at first order and we tackle this derivation by a perturbation approach. We are looking to find the perturbation $\delta\omega$ in the decomposition $\omega\approx\omega_0+\delta\omega$. A Taylor expansion of the variables $S$ and $D$ to first order and recalling $\Omega_s / \omega\approx 0$ yields
\begin{align}
    \frac{\delta\omega}{\omega_{p0}} = \pm\frac{1}{2}\gamma_0^{-2}\mu^{-1}\frac{\bar{\sigma}_0^{1/2}}{1+2\mathcal{M}}\,.\label{eq:ApproxDeltaOmega}
\end{align}
Second, we explore the limit $\omega\ll\omega_0$ as defined in Equation~(\ref{eq:Omega0}). With $\omega_0^2/\omega^2\gg 1$, Equation~(\ref{eq:Sapprox1}) is approximated as
\begin{align}
    S& \approx - \sum_s\frac{\omega_{s}^2}{\omega^2}\,.\label{eq:Sapprox1A}
\end{align}
It is now straightforward to solve for the frequency and to recover
\begin{align}
    \frac{\omega}{\omega_{p0}}=\gamma_0^{-2}\mu^{-1}\frac{\bar{\sigma}_0^{1/2}}{1+2\mathcal{M}}=2\frac{|\delta\omega|}{\omega_{p0}}\equiv \frac{\omega_\pm}{\omega_{p0}}\,.\label{eq:ApproxOmega}
\end{align}
Leptons are coupled to heavy ions via an oscillation of low frequency (compared to $\omega_0$). We derived the separation between species during this oscillation in a semi-analytic analysis for a constant radiative force ($\tilde{\mathcal{F}}_{\rm rad}=\text{const.}$) and a finite mass ratio $\mu$. The mean amplitude of the lepton oscillation (around the slowly decelerating ions) was also obtained by \citet[][Section~5]{Vanthieghem2022} and can be rewritten in the radiative units:
\begin{align}
\begin{split}
    \mathcal{A}_\pm&=\left(1+2\mathcal{M}\right)\left|\mathcal{F}_{\rm rad}\right|\;\bar{\sigma}_0^{-1/2}\\
    &=\left(1+2\mathcal{M}\right)\tilde{\mathcal{F}}_{\rm rad}\;\bar{\sigma}_0^{-1/2}\mu^{1/2}\chi^{-1/2}\gamma_{u}^{1/2}\\
    &=\left(1+2\mathcal{M}\right)\tilde{\mathcal{F}}_{\rm rad}\;\tilde{n}_{p}^{1/2}\left(\frac{\tilde{B}_{u}}{\tilde{B}_z}\right)\frac{\bar{\rho}}{\lambda}\,.
\end{split}\label{eq:approxamp}
\end{align}
Figure~\ref{fig:Validation} compares Equations~(\ref{eq:ApproxOmega}) and~(\ref{eq:approxamp}) to direct integration of the underlying equations for the basic cold plasma setup in Section~\ref{sec:couplingomega}, with varying magnetizations $\bar{\sigma}_{u}$. We find an excellent agreement between the linear theory estimates and measurements in a multi-fluid integration.

\section{Implicit-explicit time stepping}
\label{app:imexstepping}

The various time scales resolved by Equations~(\ref{eq:momentumions}) to~(\ref{eq:couplingIII}) can differ by orders of magnitude, specifically the inverse of the fast electromagnetic (plasma) frequency and the long radiative time scale $\bar{\lambda}/\bar{d}=\gamma_{u}\chi\mu^{1/2}\approx 10^{12}$. Thus, multi-fluid systems have tight constraints on the integration time step. This issue is well-known throughout the literature \citep[e.g.,][]{Amano2016,Balsara2016}, and we address it by treating the high-frequency ($\mathcal{H}$) components implicitly:
\begin{align}
    \partial_{\tilde{t}} u^x_{p} &=\mu\left(\tilde{E}_x+\beta^y_{p} \tilde{B}_z\right)\,,\label{eq:fastfirst}\\
    \partial_{\tilde{t}}\left(h_\pm u^x_\pm\right)&=\pm\left(\tilde{E}_x+\beta^y_\pm \tilde{B}_z\right)\,,\\
    \partial_{\tilde{t}} u^y_{p}&=\mu\left(\tilde{E}_y-\beta^x_{p} \tilde{B}_z\right)\,,\\
   \partial_{\tilde{t}}\left(h_\pm u^y_\pm\right)&=\pm\left(\tilde{E}_y-\beta^x_\pm \tilde{B}_z\right)\,,\\
   \partial_{\tilde{t}}\tilde{E}_x&=-\chi\left[\tilde{n}_{p} \beta^x_{p}+\tilde{n}_+\beta^x_+ -\tilde{n}_-\beta^x_-\right]\,,\\
    \partial_{\tilde{t}}\tilde{E}_y&=-\chi\left[\tilde{n}_{p} \beta^y_{p}+\tilde{n}_+\beta^y_+ -\tilde{n}_-\beta^y_-\right]\label{eq:fastlast}\,.
\end{align}
First, we write Equations~(\ref{eq:fastfirst}) to~(\ref{eq:fastlast}) as a second order Crank-Nicolson scheme of the form
\begin{align}
    \mathbf{f}_i=\mathbf{v}^{n}_i-\mathbf{v}^{n+1}_i+\frac{\Delta t}{2}\left[\mathbf{S}_i\left(\mathbf{v}^{n}\right)+\mathbf{S}_i\left(\mathbf{v}^{n+1}\right)\right]\,,
\end{align}
where $n$ denotes a specific time step and $i$ is a mesh index of the 1D domain. Furthermore, we use
\begin{align}
    \mathbf{v}=\left\{u^x_{p},u^x_+,u^x_-,u^y_{p},u^y_+,u^y_-,E_x,E_y\right\},
\end{align}
and an appropriately chosen vector of source terms $\mathbf{S}_i$. For each $i$, we solve the system $\mathbf{f}_i=\mathbf{0}$ by optimizing the vector $\mathbf{v}^{n+1}_i$ with an iterative Newton-Raphson method \citep[cf. Section~9.6 in][]{Press2007}. The low-frequency ($\mathcal{L}$) components of the system are given by Equations~(\ref{eq:momentumions}) to~(\ref{eq:photonmoments2}), as well as:
\begin{align}
    \left(\partial_{\tilde{t}}+\beta^x_{p}\partial_{\tilde{x}}\right) u^x_{p} &=\mu\tilde{g}\,,\label{eq:protonmomentumAPP}\\
    \left(\partial_{\tilde{t}}+\beta^x_\pm\partial_{\tilde{x}}\right)\left(h_\pm u^x_\pm\right)&=-2\tilde{\sigma}_\pm h_\pm u^x_\pm \tilde{n}_{\gamma\shortto u}+\tilde{g}\,,\\
    \partial_{\tilde{t}}\tilde{E}_y&=-\partial_{\tilde x}\tilde{B}_z\,,\\
    \partial_{\tilde{t}}\tilde{B}_z&=-\partial_{\tilde x}\tilde{E}_y\,.
\end{align}
We integrate the low-frequency system $\mathcal{L}$ with an explicit iterative Crank-Nicolson method \citep{Teukolsky2000}. The full decomposed system (implicit high-frequency $\mathcal{H}$ versus explicit low-frequency $\mathcal{L}$) is assembled by an operator splitting technique \citep{Strang1968,LeVeque2007} for each time step, such that 
\begin{align}
    \mathbf{v}_i^{n+1}=\exp\left[\frac{\Delta t}{2}\mathcal{H}\right]\cdot\exp\left[\Delta t\mathcal{L}\right]\cdot\exp\left[\frac{\Delta t}{2}\mathcal{H}\right]\cdot\mathbf{v}_i^{n}\,.
\end{align}
The benefits of this implicit-explicit multi-fluid integration are best exemplified by modeling the plasma oscillation of a standing Langmuir wave in a pair plasma. For time steps with $\omega_{\pm 0}\Delta t>1$, our method conserves the total energy of the system, while losing the exact phase of the oscillation. As expected, numerical diffusion errors are substituted by inaccuracies in the dispersive properties of the solver. To increase the numerical stability of the finite-difference scheme we include a hyperdiffusivity term proportional to the fourth derivative, $\mathcal{D}/\Delta t\times f^{(4)}(x)$, to the momentum equations~(\ref{eq:protonmomentum}) to~(\ref{eq:pairymomentum2}). For all the results shown throughout this work, we employ $\varepsilon=10^{-2}$.

\section{Calibration and convergence}
\label{app:calibrationB0}

\begin{figure}
  \centering
  \includegraphics[width=0.475\textwidth]{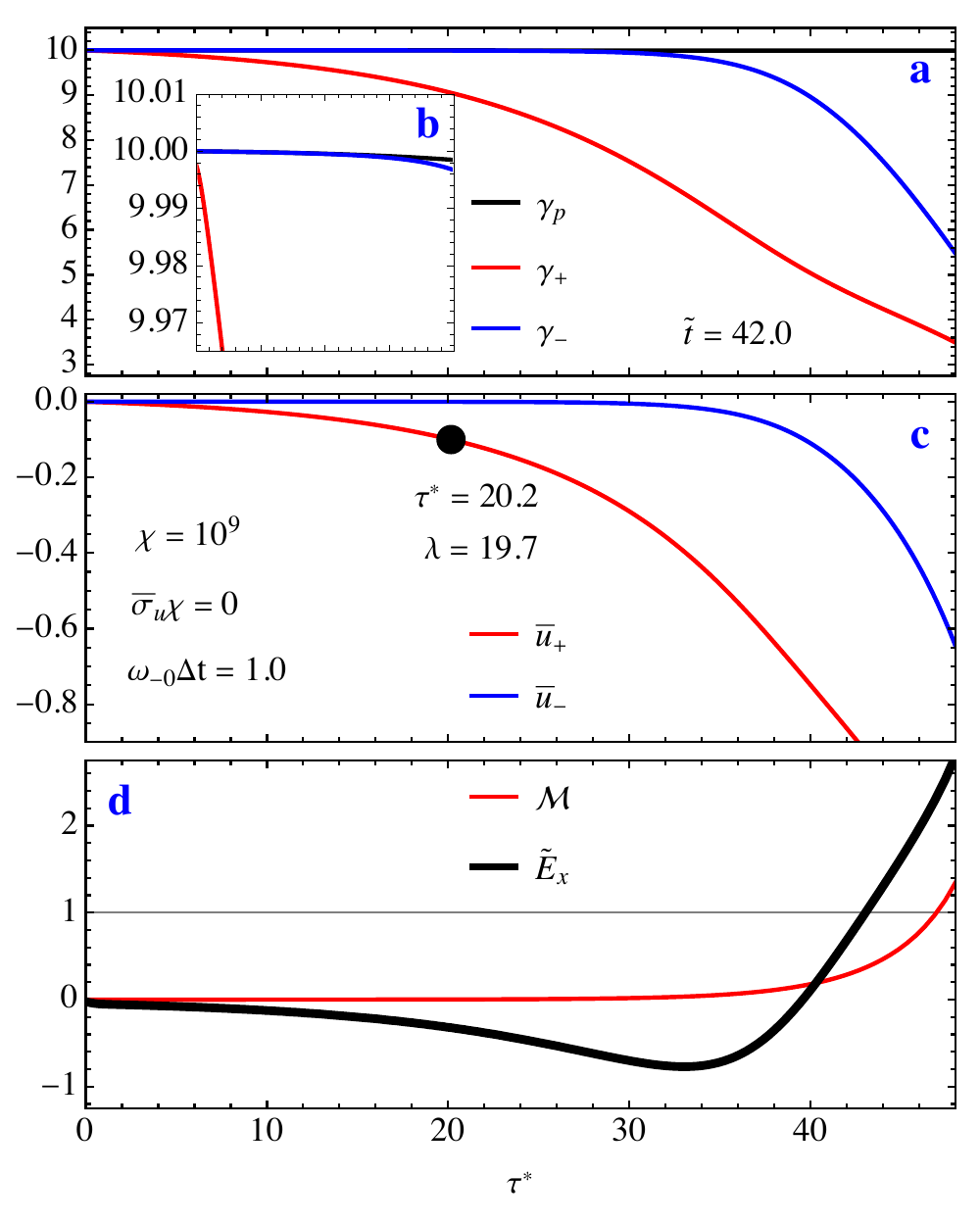}
  \vspace{-6pt}
  \caption{As Figure~\ref{fig:solutionmagnetized}, but for \emph{fully uncoupled} conditions in the unmagnetized case \citep[see][]{levinson2020b} at the time $\Bar{t}=42.0$. In the bottom panel, we show the electric field in addition to the multiplicity for an implicit-explicit integration. The inset panel (b) extends over the interval $\tau^*\in\left[0,60\right]$.}
\label{fig:solutionunmagnetized}
\end{figure}

\begin{figure}
  \centering
  \includegraphics[width=0.475\textwidth]{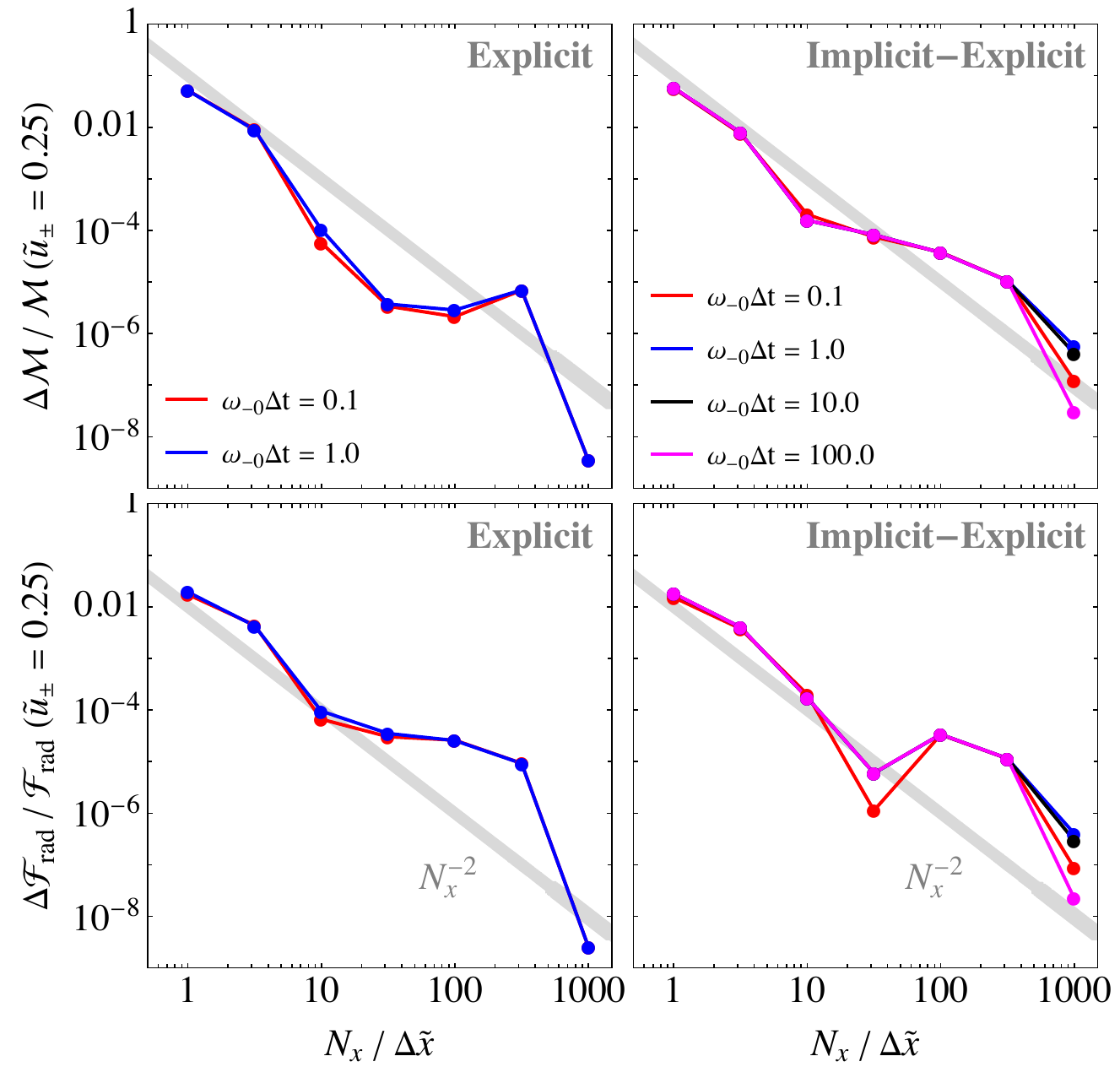}
  \vspace{-6pt}
  \caption{Convergence test for the \emph{unmagnetized} reference case shown in Figure~\ref{fig:solutionunmagnetized}. We show the relative errors of the multiplicity $\Delta\mathcal{M}/\mathcal{M}$ and radiative force $\Delta\mathcal{F}_{\rm rad}/\mathcal{F}_{\rm rad}$ measured at a velocity separation of $\bar{u}_\pm = 0.25$ for varying temporal and spatial resolutions. We compare the explicit integration (left column) to an implicit-explicit time-stepping (right column, cf.~Appendix~\ref{app:imexstepping}). The gray line scales in second order.}
\label{fig:convergenceunmagnetized}
\end{figure}

\begin{figure}
  \centering
  \includegraphics[width=0.475\textwidth]{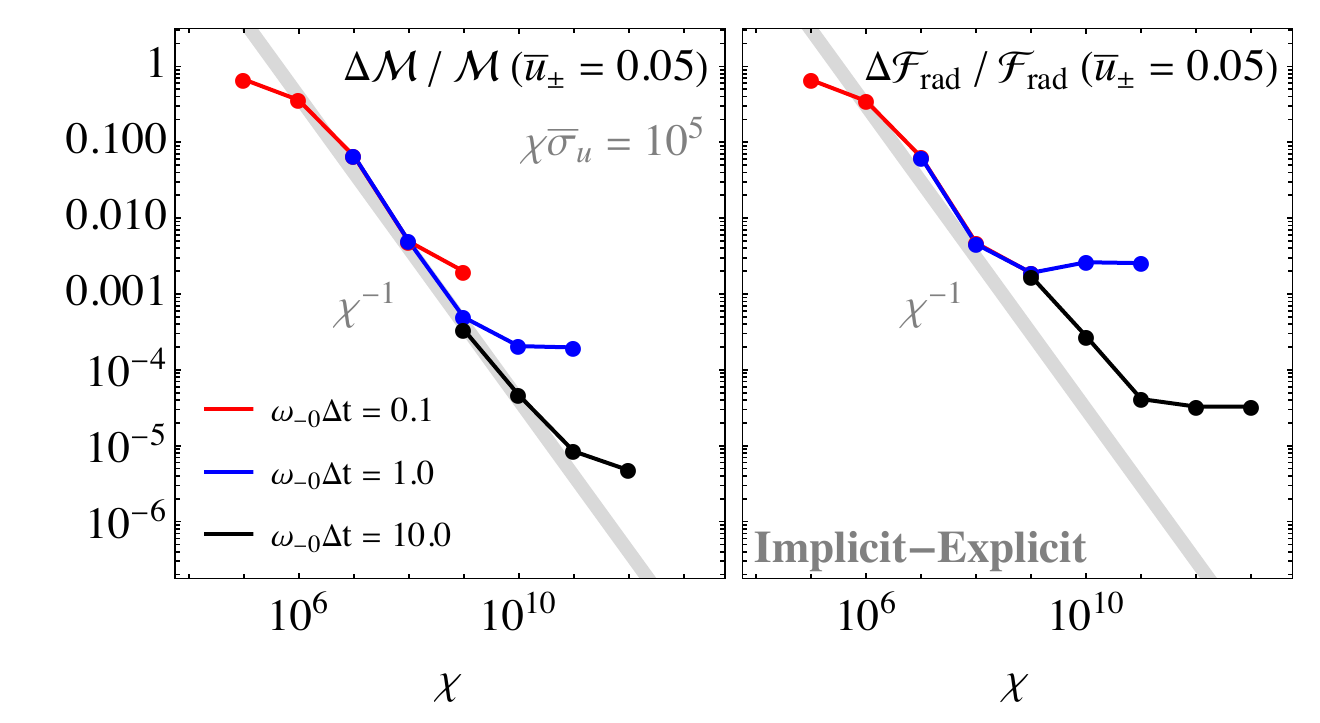}
  \vspace{-6pt}
  \caption{Relative errors of the multiplicity $\Delta\mathcal{M}/\mathcal{M}$ and radiative force $\Delta\mathcal{F}_{\rm rad}/\mathcal{F}_{\rm rad}$ for the implicit-explicit integration, measured at a velocity separation of $\bar{u}_\pm = 0.05$. We vary the scale separation in the magnetized case with a constant product $\chi\bar{\sigma}_u=10^5$. The gray line scales with first order.}
\label{fig:convergencemagnetized}
\end{figure}

We benchmark the method described in Sections~\ref{sec:1DModeling} and Appendix~\ref{app:imexstepping} against independently derived results with $\bar{\sigma}_u = 0$  \citep[Section~3.B,][]{levinson2020b}. As in Section~\ref{sec:1DEquilibrium}, we choose the seed photon density of $\tilde{n}_{\gamma\shortto u}\left(x_{u}\right)=10^{-2}$ and $\gamma_{u} = 10$. Throughout this set of tests, we vary the numerical parameter $\chi\in\left[10^3,10^{15}\right]$ and the resolution $\Delta x$ to probe the convergence of our method. Specifically, we change the number of grid points $N_x$ per lab-frame radiation length scale $\lambda$, where $N_x/\lambda\in\left[1,1000\right]$. With the CFL factor \citep{Courant1928} $f_{\rm CFL}=0.1$, and the upstream electron plasma frequency $\omega_{-0}$ (adapting Equation~\ref{eq:skindepth}) the plasma scale is resolved with $\omega_{-0}\Delta t\in\left\{0.1,1,10,100\right\}$. First, Figure~\ref{fig:solutionunmagnetized} shows an exemplary solution for the enthalpy closure and boundaries according to \citet{levinson2020b}. Our results are directly comparable to their Figure~1, where we note a difference of factor two on the $x$-axis due to another definition of $\tau^*$. The analysis presented in this paper relies on relations between the multiplicity $\mathcal{M}$ and the radiation force $\mathcal{F}_{\rm rad}$ for a specific velocity separation $\bar{u}_\pm$ between ions and pairs. As a second test, we measure the convergence of these observables with resolution in Figure~\ref{fig:convergenceunmagnetized} by displaying the difference of individual solutions to the converged result (assumed as the average between the solutions for the highest resolution reference cases). Overall, the method produces an approximate scaling with second order (gray lines in Figure~\ref{fig:convergenceunmagnetized}), while some features emerge for high resolutions. The implicit-explicit time-stepping with the operator splitting introduced in Appendix~\ref{app:imexstepping} shows convergence properties that are comparable to a fully explicit integration. Notably, the implicit-explicit scheme is capable of overstepping the time scale set by the upstream electron plasma frequency $\omega_{-0}$ by several orders of magnitude. Our method passes this convergence and, especially, is capable of producing accurate results while implicitly under-resolving the plasma frequency.

Adding a background magnetization introduces the additional scale of the Larmor radius (see Equation~\ref{eq:larmorradius}). With the product $\bar{\sigma}_u\chi$ as well as the remaining upstream conditions fixed, both the radiative length scale $\bar{\lambda}$ and the Larmor radius $\bar{\rho}$ are determined. A variation of the scale separation factor $\chi$ and the resolution $\omega_{-0}\Delta t$ then merely changes the number of grid points (or resolved skin depths) per radiative/electromagnetic length. In Figure~\ref{fig:convergencemagnetized} we demonstrate the convergence of the key observables ($\mathcal{M}$, $\mathcal{F}_{\rm rad}$) for varying resolutions and linearly increasing $\chi$. For marginally resolved plasma scales ($\omega_{-0}\Delta t=1-10$), the stationary solutions converge for fixed $\bar{\sigma}_u\chi$ with decreasing scale separation. Already for the numerically accessible regime of $\chi\sim 10^{10}$, the presented solutions are well converged with $\Delta\mathcal{M}/\mathcal{M}\lesssim 10^{-3}$. More importantly, our results suggest that the stationary solutions can be modeled with significant scale separation and then extrapolated to realistic values of $\chi$ and $\bar{\sigma}_u$ while keeping $\bar{\sigma}_u\chi$ constant. We note that the accuracy of our method for magnetized setups deteriorates for $\omega_{-0}\Delta t\gg 10$.

\section{Local plasma quantities}
\label{eq:FradNorm}

\subsection{General plasma characteristics}

In an arbitrary frame boosted along the $x$-direction with a velocity $\gamma$, we can define the magnetization as \begin{align}
\sigma_{\gamma}=\left(\frac{1}{\tilde{n}_{p}}\right)\left(\frac{B_z}{B_{u}}\right)^2\bar{\sigma}_{u}\,.
\end{align}
As before, the magnetic field $B_z$ is measured in the \emph{shock-front frame}. Equally, the Larmor radius can be written for a general boost:
\begin{align}
    \rho_\gamma=\left(\frac{\gamma}{\gamma_{u}}\right)\left(\frac{B_{u}}{B_z}\right)\bar{\rho}\,.
\end{align}
The plasma skin depth can be re-scaled as follows:
\begin{align}
    d_\gamma=\left(\frac{\gamma}{\gamma_{u}}\right)\left(\frac{1}{\tilde{n}_{p}}\right)^{1/2}\bar{d}\,.
\end{align}

\subsection{Units of the radiative force}

To calculate the radiative force acting on leptons in commonly used units \citep[$m_{p}\bar{\omega}_{p0}c$, see][]{Vanthieghem2022}, we write Equation~(\ref{eq:upmmomentum}) without normalizations:
\begin{align}
\begin{split}
&\frac{\lambda}{m_e c^2}\left(\partial_{t}+v^x_\pm\partial_{x}\right)\left(h_\pm m_e c u^x_\pm\right)\\
&=\pm\frac{\lambda e}{m_e c^2}\left(E_x+\beta^y_\pm B_z\right)-2\tilde{\sigma}_\pm h_\pm u^x_\pm \tilde{n}_{\gamma\shortto u}\,.
\end{split}
\end{align}
This expression can be recast into the following familiar form,
\begin{align}
    \frac{\text{d}\left(h_\pm m_e c u^x_\pm\right)}{\text{d}t}=\pm e\left(E_x+\beta^y_\pm B_z\right)-\frac{m_e c^2}{\lambda}2\tilde{\sigma}_\pm h_\pm u^x_\pm \tilde{n}_{\gamma\shortto u}\,,
\end{align}
where we can identify the radiative force (using Equation~\ref{eq:skindepth}):
\begin{align}
\begin{split}
    f_{\pm\rm ,rad}&=\frac{m_e c^2}{\lambda}2\tilde{\sigma}_\pm h_\pm u^x_\pm \tilde{n}_{\gamma\shortto u}\\
    &=\frac{m_e c \bar{\omega}_{p0} \gamma_{u}^{1/2}}{\chi^{1/2}\mu^{1/2}}2\tilde{\sigma}_\pm h_\pm u^x_\pm \tilde{n}_{\gamma\shortto u}\\
    &= 2\tilde{\sigma}_\pm h_\pm u^x_\pm \tilde{n}_{\gamma\shortto u}\mu^{1/2}\chi^{-1/2}\gamma_{u}^{1/2}m_{p}\bar{\omega}_{p0}c\,.
\end{split}
\end{align}

\bsp	
\label{lastpage}
\end{document}